\UseRawInputEncoding 

\documentclass[aps,showpacs,onecolumn,notitlepage,floatfix,amsmath,amssymb,nofootinbib]{revtex4-1}

\usepackage{amssymb}
\usepackage{amsmath}
\usepackage{epsfig}
\usepackage{graphicx}
\usepackage{subfig}
\usepackage{color}
\usepackage{bigints}
\usepackage{latexsym}
\usepackage{bm,latexsym}
\usepackage{mathrsfs}
\usepackage{xfrac,bigints}
\usepackage{ulem}
\usepackage{float}
\usepackage{pbox}
\usepackage[usenames,dvipsnames]{xcolor}
\definecolor{orange}{cmyk}{0,0.5,1,0}

\selectfont{}

\begin{document}

\title{ Traversable wormholes with electric and magnetic charges in general relativity theory}

\author{Pedro Ca\~nate} 
\email[]{pcannate@gmail.com }
\affiliation{Programa de F\'isica, Facultad de Ciencias Exactas y Naturales, Universidad Surcolombiana, Avenida Pastrana Borrero - Carrera 1, A.A. 385, C.P. 410001, Neiva, Huila, Colombia }

\begin{abstract}
In this work, static and spherically symmetric solutions of the general relativity coupled to linear/nonlinear electrodynamics and a dust fluid (GR-ED-DF) are studied. We demonstrate that
these solutions can be categorized into two sets, both having an electromagnetically charged metric, but with the following conditions: (i)  a variable redshift function, without a dust fluid as source, and (ii) a  constant redshift function, with a dust fluid as source. Using  (i), we provide a simple proof of the nonexistence of static and spherically symmetric traversable Morris-Thorne wormhole solutions with variable redshift functions in GR-ED-DF. Whereas using (ii), we construct several static and spherically symmetric traversable Morris-Thorne wormholes with constant redshift function in GR-ED-DF, where the source of gravity consists of a dust fluid  having negative energy density and an electromagnetic field described by a physically reasonable model of linear/nonlinear electrodynamics  
with Lagrangian density $\mathcal{L}(\mathcal{F})$ depending only on the electromagnetic invariant $\mathcal{F}\!=\!F_{\alpha\beta}F^{\alpha\beta}\!/4$, where $F_{\alpha\beta}$ are the components of the electromagnetic field tensor. We show that in the limit of the weak electromagnetic field, each of our solutions become a traversable wormhole of the electromagnetically charged Ellis-Bronnikov wormhole type.
Additionally, we present a theorem that establishes when an electrically and magnetically charged
Ellis-Bronnikov wormhole, with null $\mathcal{F}$,  can be supported by a GR-ED-DF theory.
\end{abstract}



\maketitle

\section{Introduction} 
Traversable Morris-Thorne wormholes (T-WHs) are hypothetical tunnel-like structures in spacetime
that links two distinct regions of the same spacetime (or two different spacetimes)  
and such that a subluminal particle\footnote{ That is, a particles that travel with a speed less than or equal to the speed of light $v\leq c$. }  could travel from one ``end of the tunnel" to the other and return back  (see \cite{morris88,morris88-2,visser95} for details).  
Like black holes, the concept of traversable wormhole (T-WH) was first introduced in the context of general relativity (GR), but in contrast to black hole solutions, T-WHs are subject to some severe restrictions.
The main restriction is the fact that in the context of GR 
the construction of T-WHs necessarily demands the presence of a type of matter 
with negative energy density and whose energy momentum tensor violates the null energy condition (NEC),
leading to what is called ``exotic matter" in Refs. \cite{morris88,morris88-2}. \\
One of the first, and simplest examples of wormholes in general relativity is the ultrastatic and spherically symmetric Ellis-Bronnikov wormhole metric (discovered in 1973 by Ellis \cite{ellis73} and, independently, Bronnikov \cite{Bronnikov73}), it represents a massless wormhole which is sustained by a phantom scalar field\footnote{That is, a massless scalar field with a negative kinetic term.}. 
Later, Morris and Thorne \cite{morris88} studied this wormhole and proved it to be traversable.  
In the current literature, specifically in Refs. \cite{Shatskii08,Bronnikov13}, it was shown that 
the Ellis-Bronnikov wormhole metric is also a solution to the Einstein equations 
with a material source represented as a superposition of a spherically symmetric magnetic (or electric) field and negative-density dust matter, serving as exotic matter necessary for a traversable wormhole to exist. 
This solution has been extensively studied, and its properties like gravitational lensing, quasinormal modes, shadows and stability have all been thoroughly investigated \cite{EllisLensing,QNMEllis,EllisShadows,EllisStability}. 

The investigation of traversable wormhole spacetimes (T-WHS) is currently generating a lot of interest, as certain properties commonly associated with black holes can be closely mimicked by traversable wormholes \cite{Biswas22,Franzin22,Volkel18,Cardoso16,Cardoso16B}.
It is worth noting that massless wormholes are particularly interesting because they are expected to have unique gravitational lensing effects \cite{Abe10,Ishkaeva23}, moreover, current research findings suggest that both massive and massless wormholes can reproduce postmerger initial ring-down gravitational waves characteristic of black hole horizon \cite{Izmailov19,Konoplya16}.
In addition, stable massless wormholes are not only fascinating from a theoretical point of view, but also have important astrophysical applications, especially as galactic halo objects \cite{Lukmanova18}. 

Since obtaining the Ellis-Bronnikov wormhole metric, several new traversable wormholes have been presented as exact solutions of general relativity. Some recent examples are; T-WHs with electric or magnetic charges in the Einstein-Maxwell-dilaton theory  
\cite{Clement09,Goulart17,Huang19};  
Traversable wormhole solutions of Einstein-nonlinear electrodynamics gravity coupled to a self-interacting phantom scalar field \cite{Canate2022}; T-WHs  
that are in general not spherically symmetric \cite{Bronnikov022,Visser89}; and for whom many types of exotic matter fields are introduced to achieve the traversability of wormholes.

Recently, in order to avoid the need to use exotic matter in the construction of T-WHs, several papers discussed the existence of T-WHs in modified gravity theories\footnote{Modified gravity theories have received increased attention lately due to combined motivation coming from high-energy physics, cosmology and astrophysics.} (for examples, see Refs. \cite{Canate2022,duplessis15,nascimento17,lobo08,zangeneh15,ovgun19,zubair18,bohemer12,evseev18,mironov18,Harko14,WH_5EGB,Kanti2011,antoniou19,kanti11,Canate2019,canate19b}).  
More precisely, the necessity of the presence of exotic matter in T-WHs can be circumvented in modified theories of gravity which contain higher curvature corrections in their gravitational actions, so that additional curvature degrees of freedom can support such geometries.
Another alternative for the requirement of exotic matter in the traversable wormholes, 
consists to use quantum matter fields in frameworks of semi-classical quantum gravity.
The point is that the quantum fields can easily violate NEC \cite{Epstein65}.
From this perspective, recently it has been shown that quantum matter fields can provide the necessary 
amount of negative energy density to create the semiclassical traversable wormholes (for examples, see Refs. \cite{Blazquez21,Konoplya22,gao17,fu19,horowitz19}).  

In this work, we will leave aside the issue of exotic matter in the T-WHs.
Instead, the main interest of this paper is the construction of electric and/or magnetic generalizations\footnote{ I.e., the solution contains at least one electric or magnetic field.}
of the Ellis-Bronnikov wormhole in the framework of general relativity coupled to (linear or nonlinear) electrodynamics and an exotic dust matter ({\it i.e.,} a pressureless perfect fluid with negative energy density). 
In our case the exotic dust fluid is solely responsible for the violation of the null-energy condition required for the traversability of wormhole. Whereas, the T-WH geometry cannot have a nonzero shape function without a linear (or nonlinear) electrodynamics source.  

The nonlinear electrodynamics (NLED) theories are extensions of Maxwell's electromagnetism [or linear electrodynamics (LED)] whose Lagrangian density of the electromagnetic field depends in a nonlinear way on the two electromagnetic invariants, $\mathcal{F}= 2(\boldsymbol{\mathcal{B}}^2-\boldsymbol{\mathcal{E}}^2)$ and $\mathcal{G} = \boldsymbol{\mathcal{E}}\cdot\boldsymbol{\mathcal{B}}$,  where $\boldsymbol{\mathcal{E}}$ and $\boldsymbol{\mathcal{B}}$ are respectively the electric and magnetic fields, and hence, an arbitrary NLED Lagrangian density is characterized by a function $\mathcal{L}(\mathcal{F}, \mathcal{G})$ of the electromagnetic invariants, $\mathcal{F}$ and $\mathcal{G}$.  
For more details on these aspects see \cite{BI,JFP70,Salazar87,Heuler_35,Heisenberg_36} (see also \cite{Sorokin21} for a recent review) and references within.
It is notable that the several NLED models have been extensively studied as possible guides
to catch new physics, for instance: for tackling fundamental cosmological problems (dark energy \cite{Labun10,RationalNLED}  and dark mater \cite{Beltran21}); as a solution of the black hole singularity problem (see Refs. \cite{Ayon1998,Bronnikov_Ayon1998,Bronnikov2000,canate2023}); and in the context of AdS/CFT correspondence and gravity/CMT holography (see Refs. \cite{Sorokin21,ExpNLED,Jing11}). 

Let us proceed by introducing a few noteworthy $\mathcal{L}(\mathcal{F}, \mathcal{G})$ model examples.
Naturally, the most simple $\mathcal{L}(\mathcal{F},\mathcal{G})$ model corresponds to the linear case (LED),  
which is provided by, 
\begin{equation}\label{Maxw}
\mathcal{L}_{_{_{\mathrm{LED}}}} = \mathcal{F} 
\end{equation}
which is known as Maxwell's theory of electrodynamics.\footnote{
The  Maxwell's electrodynamics is one of the most notable and experimentally verified classical field theories 
ever constructed.  Since its formulation (about 1860), it has been the source of remarkable predictions such as the electromagnetic radiation. In addition, the Maxwell's theory has served as a keystone for the proposal of new theories, such as Einstein's theory of special relativity.}

It can be stated that, the first nonlinear generalization of Maxwell's electrodynamics for strong fields was constructed by Born and Infeld (BI) in 1933-1934. 
Specifically, the BI electrodynamics was invented to ensure that electric field self-energy of charged point particles is finite and, therefore, a solution of the Maxwell's electrodynamics problem of point charges and their diverging self-energy is proposed \cite{BI}.
The proposed BI Lagrangian that depends nonlinearly on the electromagnetic invariants $\mathcal{F}$ and $\mathcal{G}$ was inspired in a finiteness principle for the electromagnetic field magnitude (analogous to the special relativity theory that assumed an upper limit to the velocity of light), and has the form
\begin{equation}\label{BI2}
\mathcal{L}_{_{\mathrm{BI}}}  
= 4\beta^{2} \left( -1 + \sqrt{ 1 + \frac{\mathcal{F}}{2\beta^{2}} + \frac{\mathcal{G}^{2}}{16\beta^{4}} 
}~\right),
\end{equation}
where $\beta$ is a constant which has the physical interpretation of a critical field strength. 

Later, in 1935, a new NLED model was formulated by Euler and Heisenberg (EH)
which computed a complete effective action describing nonlinear corrections to Maxwell's theory due to quantum electron-positron one-loop effects. In a nonperturbative form, it is given by
\begin{equation}\label{EH}
\mathcal{L}_{_{\mathrm{EH}}}  
= \mathcal{L}_{_{_{\mathrm{LED}}}} + \frac{1}{8\pi^{2}}\int^{^{\infty}}_{_{0}} \frac{ e^{-m^{2}_{e}\varrho} }{\varrho^{3}}\left[ \left(q_{e}\varrho\right)^{2}\frac{ {\tt Re}\!\left\{ \cosh\left( q_{e}\varrho \sqrt{ 2\mathcal{F}+ 2i\mathcal{G} }  \right) \right\} }{{\tt Im}\!\left\{ \cosh\left( q_{e}\varrho \sqrt{ 2\mathcal{F}+ 2i\mathcal{G} }  \right) \right\}}\mathcal{G} - \frac{2}{3}\left(q_{e}\varrho\right)^{2}\mathcal{F} - 1 \right] d\varrho
\end{equation}
where $m_{e}$ is the mass of the electron and $q_{e}$ is the elementary charge.  
Writing $\mathcal{L}_{_{\mathrm{EH}}}$ 
as a power series of $\mathcal{F}$ and $\mathcal{G}$ yields 
\begin{equation}\label{EH1}
\mathcal{L}_{_{\mathrm{EH}}} = \mathcal{L}_{_{_{\mathrm{LED}}}} - \frac{16\alpha^{2}}{45m^{4}_{e}}\left( \frac{\mathcal{F}^{2}}{2} + \frac{7}{8}\mathcal{G}^{2}\right) + \mathcal{O}\!\left(\mathcal{F}^{3},\mathcal{G}^{3},\mathcal{F}^{2}\mathcal{G},\mathcal{F}\mathcal{G}^{2}\right)
\end{equation}
where $\alpha=q^{2}_{e}/(4\pi)$ is the fine structure constant.
Assuming that the electromagnetic field is sufficiently small, and taking into account the terms up to the quadratic order of $\mathcal{F}$ and $\mathcal{G}$, the EH Lagrangian is approximated
by
\begin{equation}\label{EH2}
\mathcal{L}_{_{\mathrm{EH}}} = \mathcal{L}_{_{_{\mathrm{LED}}}}  - \frac{\mu^{2}}{2}\mathcal{F}^{2}
- \frac{7\mu^{2}}{8}\mathcal{G}^{2}
\end{equation}
which corresponds to the weak field approximation of \cite{Heuler_35,Heisenberg_36},  and the coupling constant $\mu^{2}$ is written as $\mu^{2}=\frac{16\alpha^{2}}{45m^{4}_{e}}$.  
In addition to Born-Infeld and Euler-Heisenberg theories, other types of  
$\mathcal{L}(\mathcal{F}, \mathcal{G})$ electrodynamics
(containing such instances as power law \cite{Hassaine08,Gurtug12}, inverse \cite{Cembranos15,Gaete21}, exponential \cite{ExpNLED}, rational \cite{RationalNLED}, logarithmic \cite{LogNLED}, double logarithmic \cite{2LogNLED} and other NLEDs \cite{wormholes}) have been discussed in the literature.
They  have been created for applications in gravity and cosmology, as well as for a gravity/CMT holographic description of certain strongly coupled condensed matter systems \cite{Sorokin21}.

In the next section we first review the basics of traversable wormholes. Second, the field equations for general relativity coupled to (linear or nonlinear) electrodynamics and a dust fluid (GR-ED-DF), are obtained. 
And third, the electric-magnetic duality of static and spherically symmetric solutions in GR-ED-DF is presented.
Section \ref{secIII} contains a theorem on the nonexistence of static and spherically
symmetric traversable Morris-Thorne wormholes with nontrivial redshift functions\footnote{  I.e., variable redshift functions.} in GR-ED-DF. 
In contrast, Sec. \ref{secIV} concludes that the only permissible static and spherically symmetric T-WHs within this gravity framework are those with a trivial redshift functions\footnote{ I.e., constant redshift function.}. 

Some examples of electromagnetically charged\footnote{I.e., at least one nonzero charge that is either electrical or magnetic is present.} 
T-WHs with trivial redshift functions in general relativity supported by (linear or nonlinear) electromagnetic fields and exotic dust fluids are displayed.  
Our conclusions and final remarks are presented in Sec. \ref{secV}. 
In this paper we use units where $G = c = \hbar =\epsilon_{_{0}}=\mu_{_{0}} =1$, and the metric signature $(-+++)$. Greek indices run over all four spacetime coordinates, whereas Latin indices only run over the three spatial coordinates.

\section{Basics on traversable wormholes, field equations for GR-ED-DF, and electric-magnetic duality of static and spherically symmetric solutions in GR-ED-DF}\label{UTW_GR-NLED-DF}

Our first task in this section is to review the basic features of traversable Morris-Thorne wormholes. Later, the field equations of general relativity coupled to (linear or nonlinear) electrodynamics and a dust fluid are determined for a static, spherically symmetric, electrically and magnetically charged spacetime. The final point is to introduce the electric-magnetic duality of static and spherically symmetric solutions in GR-ED-DF.

\subsection{Traversable wormhole spacetimes of the Morris-Thorne type}
In accordance with \cite{morris88,morris88-2}, the metric ansatz to describe a static, spherically symmetric and asymptotically flat traversable wormhole spacetime of the Morris-Thorne type in standard coordinates $(t,r,\theta,\varphi)$, also known as Schwarzschild like coordinates, is given by
\begin{equation}\label{SSSmet}
\boldsymbol{ds}^{2} =  - e^{ 2\Phi(r) }\boldsymbol{dt}^{2} + \frac{\boldsymbol{dr}^{2}}{1 - \frac{b(r)}{r}}  + r^{2}\boldsymbol{d\Omega}^{2},   
\end{equation}
where $\Phi(r)$ and $b(r)$ are smooth functions, respectively known as redshift function (since this gives a measure of the gravitational redshift) and shape function (since this determines the topological configuration of the spacetime). Here, to simplify the notation, we used $\boldsymbol{d\Omega}^{2} = \boldsymbol{d\theta}^{2}  \!+\! \sin^{2}\theta \boldsymbol{d\varphi}^{2}$ as the line element of a two-dimensional sphere of radius $r$.
Additionally, the wormhole spacetime is identified by the presence of a throat that is a two-sphere of radius  
radius $r_{_{0}}$ satisfying $b(r_{_{0}}) = r_{_{0}}$, acting as a membrane connecting the two sides of the wormhole. The range of the $r$ coordinate is $r\in[r_{_{0}},\infty)$, therefore, the coordinate $r$ has a special geometric interpretation such that  $4\pi r^2$ is the area of a sphere centered on the wormhole throat.
On the other hand, as was shown in Refs. \cite{morris88,morris88-2}, for the wormhole to be traversable, the fulfilling of the following conditions is required: 
\begin{eqnarray} 
&&\textup{{\it Wormhole domain:}}\hspace{2.3cm}   
 1 - \frac{b(r)}{r} > 0  \quad\quad\quad\quad \forall\quad\!\!\!\!\!r>r_{_{0}} \label{TWC1}\\[2.7pt]
&& \textup{{\it Absence of horizons:}}\hspace{2cm}  e^{2\Phi(r)}\in\mathbb{R}^{+}\!-\!\{0\}  
\quad\quad  \forall\quad\!\!\!\!\! r\geq r_{_{0}} \label{TWC2}\\[2.7pt]
&&\textup{{\it Flare-out condition:}}\hspace{2.1cm} b'(r)\Big|_{r=r_{_{0}}}<1\label{TWC3}\\
&&\textup{{\it AF spacetime:}}\hspace{3.0cm} \lim\limits_{r \to \infty}\Phi(r)=0 \quad\quad\textup{and} 
\quad\quad \lim\limits_{r \to \infty} \frac{b(r)}{r}=0 \label{AFWH}
\end{eqnarray}
where the prime $'$ denotes derivative with respect to $r$. 
For the case, $\Phi(r)=constant$, the line element Eq.\!~\eqref{SSSmet}  becomes the metric ansatz for an arbitrary ultrastatic\footnote{
A spacetime is called ultrastatic if it admits an atlas of charts in which the metric tensor takes the
form
\begin{equation}\label{UltraS}
\boldsymbol{ds^{2}} = - \boldsymbol{dt^{2}} + g_{jl}\boldsymbol{dx^{j}}\boldsymbol{dx^{l}}
\end{equation}  
where the Latin indices label spatial coordinates only, 
where $t$ is the time relative to a free-falling observer moving with four-velocity $u^{\alpha} = - g^{\alpha\beta}\partial_{\beta}t$, and where the metric coefficients $g_{jl}$ are independent of the time coordinate $t$. In an ultrastatic spacetime the only nonvanishing Christoffel symbols are $\Gamma^{i}{}_{jl}$ (and their partial derivatives with respect to spatial coordinates), see \cite{Stephani1900} for details. 
In other words, computing the Christoffel symbols for the metric \eqref{UltraS}, we get
$\Gamma^{t}{}_{\alpha\beta} = \Gamma^{\alpha}{}_{\beta t} = \partial_{_{t}}\Gamma^{i}{}_{jl} = 0$,       
which implies that the only components of the Riemann tensor $R^{\alpha}{}_{\beta\mu\nu}$, of the metric (\ref{UltraS}), that do not vanish identically, coincide with those of the Riemann tensor $^{^{(3)}}\!\!R^{i}{}_{jkl}$ of the three-dimensional metric $\boldsymbol{^{^{(3)}}\!\!ds^{2}}=g_{jl}\boldsymbol{dx^{j}}\boldsymbol{dx^{l}}$.  As a consequence, the differential geometrical properties of ultrastatic spacetime \eqref{UltraS} are completely determined by  the Riemannian metric induced on three-dimensional hypersurface $t$ = constant, 
$\boldsymbol{^{^{(3)}}\!\!ds^{2}}=g_{jl}\boldsymbol{dx^{j}}\boldsymbol{dx^{l}}$ (see  \cite{Sonego2010,Fulling,Fulling81,DonPage} for details). } 
and spherically symmetric spacetime \cite{Fulling,DonPage,Sonego2010}, 
and if in addition to this \eqref{TWC1}$-$\eqref{AFWH} are also satisfied,
then this spacetime geometry admits an ultrastatic, spherically symmetric and asymptotically flat traversable wormhole interpretation (see for details \cite{Pcanate2023}).

\subsection{ Einstein gravity coupled to (linear or nonlinear) electrodynamics and a dust fluid }
Now, we focus on obtaining the relevant motion equations of the general relativity coupled to (linear or nonlinear) electrodynamics and a dust fluid for a static and spherically symmetric spacetime geometry.   

The gravitational action for GR-ED-DF is given 
\begin{equation}\label{actionL}
S[g_{ab},A_{a},\rho_{_{D\!F}}] = \frac{1}{16\pi}\int d^{4}x \sqrt{-g} \left[  R - 4\mathcal{L}(\mathcal{F}) - \rho_{_{D\!F}}\right], 
\end{equation}
where $R$ denotes the Ricci scalar, $\mathcal{L}(\mathcal{F})$ is a function that represents the Lagrangian density for the electromagnetic field depending only on the electromagnetic invariant 
$\mathcal{F}\equiv \frac{1}{4}F_{\alpha\beta}F^{\alpha\beta}$ being $F_{\alpha\beta}=2\partial_{[\alpha}A_{\beta]}$  
the components of the electromagnetic field tensor $\boldsymbol{F}=\frac{1}{2}F_{\alpha\beta} \boldsymbol{dx^{\alpha}} \wedge \boldsymbol{dx^{\beta}}$ with $A_{\alpha}$ the components of the electromagnetic potential, whereas $\mathcal{L}_{_{D\!F}}=-\rho_{_{D\!F}}$ is the Lagrangian density for a dust fluid\footnote{Also called dust matter in the cosmological context.} being $\rho_{_{D\!F}}$ the proper energy density of the fluid (also known as the energy density of the dust in its rest frame). 
\\
Using the notation $\mathcal{L}^{n}_{_{\mathcal{F}}}= (\frac{d\mathcal{L}}{d\mathcal{F}})^{n}$, 
the GR-ED-DF field equations obtained from the variation of Eq.\!~\eqref{actionL} are
\begin{equation}\label{GR_ED_DF} 
G_{\alpha}{}^{\beta} =  8\pi (T_{\alpha}{}^{\beta})\!_{_{_{D\!F}}} + 8\pi (T_{\alpha}{}^{\beta})\!_{_{_{E\!D}}}
\end{equation}
where $G_{\alpha}{}^{\beta} = R_{\alpha}{}^{\beta} - \frac{R}{2} \delta_{\alpha}{}^{\beta}$ are the components of the Einstein tensor; $(T_{\alpha}{}^{\beta})\!_{_{_{D\!F}}} = \rho_{\!\!_{_{_{D\!F}}}} u_{\alpha}u^{\beta}$ are the components of the dust fluid energy-momentum tensor,
with energy density $\rho_{\!\!_{_{_{D\!F}}}}$ as measured by a comoving observer with four-velocity $\boldsymbol{u}=u^{\alpha}\partial_{\alpha}$ such that $\boldsymbol{ds}^{2}(\boldsymbol{u},\boldsymbol{u}) = -1$, satisfying the equations of motion
\begin{equation}\label{DF_mot_eq}
\nabla_{\beta}(T_{\alpha}{}^{\beta})\!_{_{_{D\!F}}} = 0,
\end{equation}
whereas 
\begin{equation}\label{TEM_ED}
4\pi(T_{\alpha}{}^{\beta})\!_{_{_{E\!D}}} = \mathcal{L}_{\mathcal{F}} \hskip.01cm F_{\alpha\mu}F^{\beta\mu} - \mathcal{L}\hskip.01cm\delta_{\alpha}{}^{\beta},    
\end{equation}
are the components of the linear/nonlinear electrodynamics energy-momentum tensor, with electromagnetic field tensor $F_{\mu\nu}$ fulfilling the source-free modified Maxwell's equations 
\begin{equation}\label{e1}   
\nabla_{\mu}(\mathcal{L}_{\mathcal{F}}F^{\mu\nu}) = 0 = d\boldsymbol{F}.
\end{equation}
An alternative representation for linear/nonlinear ED, the $(\mathcal{H},\mathcal{P})-$formalism, can be defined using the auxiliary antisymmetric tensor $\boldsymbol{P}=\frac{1}{2}P_{\alpha\beta} \boldsymbol{dx^{\alpha}} \wedge \boldsymbol{dx^{\beta}}$, with components
$P_{\alpha\beta} = \mathcal{L}_{\mathcal{F}} F_{\alpha\beta}$, and the electromagnetic Hamiltonian density $\mathcal{H}$ (also known as structural function) obtained from the electromagnetic Lagrangian density by means of a Legendre transformation
(see for instance \cite{Gutierrez1981,Salazar87,Salazar89}),
\begin{equation}\label{actionHP} 
\mathcal{H} = 2\mathcal{F}\mathcal{L}_{\mathcal{F}} -  \mathcal{L}. 
\end{equation}  
It can be shown that $\mathcal{H}$ is a function of the 
invariant $\mathcal{P}$, defined as 
\begin{equation}\label{def_P}
\mathcal{P} = \frac{1}{4}P_{\alpha\beta}P^{\alpha\beta} = (\mathcal{L}_{\mathcal{F}})^{\!^{2}}\mathcal{F}.    
\end{equation}
Since $d\mathcal{H} = \mathcal{H}_{\mathcal{P}}d\mathcal{P}  
= (\mathcal{L}_{\mathcal{F}})^{-1}d[(\mathcal{L}_{\mathcal{F}})^{\!^{2}}\mathcal{F}]$, where
$\mathcal{H}_{\mathcal{P}}\equiv \frac{d\mathcal{H}}{d\mathcal{P}}$, yields $\mathcal{H}_{\mathcal{P}}=(\mathcal{L}_{\mathcal{F}})^{\!^{-1}}$. 
With the help of $\mathcal{H}$ one expresses the electromagnetic Lagrangian density  in the action (\ref{actionL}) as 
\begin{equation}\label{actionLF}
\mathcal{L}=2\mathcal{P}\mathcal{H}_{\mathcal{P}} - \mathcal{H}.  
\end{equation} 
While the source-free modified Maxwell equations \eqref{e1}, in terms of $P_{\mu\nu}$ and $\mathcal{H}$, take the form
\begin{equation}\label{e1HP}
\nabla_{\mu}(P^{\mu\nu}) = 0 = d(\mathcal{H}_{\mathcal{P}}\boldsymbol{P}).
\end{equation}
Thus, in the $(\mathcal{H},\mathcal{P})-$formalism, the components $(T_{\alpha}{}^{\beta})\!_{_{_{E\!D}}}$ can be written as,
\begin{equation}\label{HP_EM_tensr}
4\pi(T_{\alpha}{}^{\beta})\!_{_{_{E\!D}}} = \mathcal{H}_{\mathcal{P}} \hskip.01cm P_{\alpha\mu}P^{\beta\mu} - (2\mathcal{P}\mathcal{H}_{\mathcal{P}} - \mathcal{H})\hskip.01cm\delta_{\alpha}{}^{\beta}.    
\end{equation}
The $(\mathcal{L},\mathcal{F})$ and $(\mathcal{H},\mathcal{P})$  formalisms of the same theory are not always equivalent. More precisely, according with \cite{Bronnikov_Ayon1998,Bronnikov2000}, a theory initially specified by $\mathcal{L}(\mathcal{F})$ is equivalently reformulated in the $\mathcal{H}(\mathcal{P})$ framework only in a range of where $\mathcal{F}(\mathcal{P})$ is a monotonic function. 
\subsubsection{GR-ED-DF  equations in static and spherically symmetric  background} 
We are now interested in describing the GR-ED-DF equations
for an arbitrary static and spherically symmetric 
T-WH geometry.
For the spacetime geometry \eqref{SSSmet}, the non-null components of the Einstein tensor are
\begin{equation}  
G_{t}{}^{t} = -\frac{b'}{r^{2}}, \quad\quad G_{r}{}^{r} = \frac{2}{r}\!\!\left(1-\frac{b}{r}\right)\!\!\Phi' - \frac{b}{r^{3}}, \quad\quad G_{\theta}{}^{\theta} = G_{\varphi}{}^{\varphi} = \left(1- \frac{b}{r}\right)\!\!\left( \Phi'^{2} + \Phi''\right) + \frac{ 2r - b - rb' }{ 2r^{2} }\Phi' - \frac{b'}{ 2r^{2} } + \frac{b}{ 2r^{3} }. 
\end{equation}
On the other hand, in this spacetime geometry, relative to a comoving observer with 4-velocity  
$\boldsymbol{u}=-e^{-\Phi}\boldsymbol{\partial_{t}}$,
the component of the energy-momentum tensor for dust fluid
takes the form
\begin{equation}\label{T_DF}
(T_{\alpha}{}^{\beta})\!_{_{_{D\!F}}} = - \rho_{\!\!_{_{_{D\!F}}}} \delta_{\alpha}{}^{t}\delta_{t}{}^{\beta},    
\end{equation}
with rest energy density depending only on the radial coordinate [{\it i.e.,} $\rho_{\!\!_{_{_{D\!F}}}} = \rho_{\!\!_{_{_{D\!F}}}}(r)$].  
Thus, for static and spherically symmetric spacetime the equations \eqref{DF_mot_eq} are trivially satisfied ({\it i.e.,} $\partial_{_{t}}\rho_{\!\!_{_{_{D\!F}}}}=0$). 
\\
%
Regarding the electromagnetic field tensor, since the spacetime metric \eqref{SSSmet} is static and spherically symmetric, the only nonvanishing terms are the electric $F_{tr}$ and magnetic $F_{\theta\phi}$ components, in such a way that
\begin{equation}\label{Fab_SSS} 
F_{\alpha\beta} = \Big( \delta^{t}_{\alpha}\delta^{r}_{\beta} - \delta^{r}_{\alpha}\delta^{t}_{\beta}\Big) F_{tr} + \Big( \delta^{\theta}_{\alpha}\delta^{\varphi}_{\beta} - \delta^{\varphi}_{\alpha}\delta^{\theta}_{\beta}\Big) F_{\theta\varphi}.       
\end{equation}
Hence, the nonvanishing components of $(T_{\alpha}{}^{\beta})\!_{_{_{E\!D}}}$
obtained from  the metric given in Eq.\!~\eqref{SSSmet}, with the electromagnetic field tensor in Eq.\!~\eqref{Fab_SSS}, and an arbitrary 
$\mathcal{L}(\mathcal{F})$, are given by,
\begin{eqnarray}
&& 8\pi (T_{t}{}^{t})\!_{_{_{ E \! D}}} = 8\pi (T_{r}{}^{r})\!_{_{_{ E \! D}}} = 2(F_{tr}F^{tr}\mathcal{L}_{\mathcal{F}}-\mathcal{L}), \label{NLEDtt}\\
&&8\pi (T_{\theta}{}^{\theta})\!_{_{_{ E \! D}}} = 8\pi (T_{\varphi}{}^{\varphi})\!_{_{_{ E \! D}}} = 2(F_{\theta\varphi}F^{\theta\varphi}\mathcal{L}_{\mathcal{F}}-\mathcal{L}). \label{NLEDtheph} 
\end{eqnarray}  %
Using Eqs.\!~\eqref{e1}, written as  $\partial_{\alpha}(\sqrt{-g}\mathcal{L}_{\mathcal{F}}F^{\alpha\beta}) = 0 = d(F_{\alpha\beta} \boldsymbol{dx^{\alpha}} \wedge \boldsymbol{dx^{\beta}})$, it follows that
\begin{equation}
F^{rt} = \frac{ q_{_{\mathcal{E}}} }{r^{2}\sqrt{-g_{tt}g_{rr}}~\mathcal{L}_{\mathcal{F}} }, \quad\quad\quad\quad\quad
F^{\theta\varphi} = \frac{q_{_{\mathcal{B}}}}{r^4 \sin\theta },
\end{equation}
where $q_{_{\mathcal{E}}}$ and $q_{_{\mathcal{B}}}$ are the electric and magnetic charges of the spacetime, respectively.
Thus, the non-null components of $F_{\alpha\beta}$  and the invariant $\mathcal{F}$ for the generic 
static and spherically symmetric spacetime geometry Eq.\!~\eqref{SSSmet}, are given by
\begin{equation}\label{F_invariant}
F_{rt} = -\frac{ \sqrt{-g_{tt}g_{rr} } ~ q_{_{\mathcal{E}}}  }{r^{2}\mathcal{L}_{\mathcal{F}}},\quad\quad\quad  F_{\theta\varphi} = q_{_{\mathcal{B}}} \sin\theta, \quad\quad\quad \mathcal{F} = \frac{1}{2r^{4}}\!\!\left(q^{2}_{_{\mathcal{B}}}  - \frac{ q^{2}_{_{\mathcal{E}}} }{ \mathcal{L}^{^{2}}_{\mathcal{F}} }\right).
\end{equation}
Hence, Eq.\!~\eqref{F_invariant} is the most general solution of Eqs.\!~\eqref{e1} for static and spherically symmetric spacetimes in the $(\mathcal{L},\mathcal{F})-$formalism.
 
%
In the $(\mathcal{H},\mathcal{P})-$formalism the non-null components of $P_{\alpha\beta}$ and the invariant $\mathcal{P}$, for the generic static and spherically symmetric geometry \eqref{SSSmet}, are respectively given by
\begin{equation}\label{P_invariant}
P_{rt} = -\frac{ \sqrt{-g_{tt}g_{rr}}~q_{_{\mathcal{E}}}  }{r^{2} },\quad\quad\quad  P_{\theta\varphi} = \frac{q_{_{\mathcal{B}}} \sin\theta}{\mathcal{H}_{\mathcal{P}}}, \quad\quad\quad
\mathcal{P} = \frac{1}{2r^{4}}\!\!\left(\frac{q^{2}_{_{\mathcal{B}}}}{\mathcal{H}^{^{2}}_{\mathcal{P}}}  - q^{2}_{_{\mathcal{E}}}  
\right).
\end{equation}
Thus, Eq.\!~\eqref{P_invariant} represents the general solution of Eqs.\!~\eqref{e1HP} for a static and spherically symmetric spacetime, while the nonvanishing components of $(T_{\alpha}{}^{\beta})\!_{_{_{ E \! D}}}$ in the $(\mathcal{H},\mathcal{P})-$formalism are given by
\begin{eqnarray}
&& 8\pi (T_{t}{}^{t})\!_{_{_{ E \! D}}} = 8\pi (T_{r}{}^{r})\!_{_{_{ E \! D}}} 
= 2(\mathcal{H} -P_{\theta\varphi}P^{\theta\varphi}\mathcal{H}_{\mathcal{P}}), \label{NLEDttHP}\\
&&8\pi (T_{\theta}{}^{\theta})\!_{_{_{ E \! D}}} = 8\pi (T_{\varphi}{}^{\varphi})\!_{_{_{ E \! D}}} 
=2(\mathcal{H}-P_{tr}P^{tr}\mathcal{H}_{\mathcal{P}}). \label{NLEDththHP} 
\end{eqnarray}  
Next we shall use these expressions to solve the GR-ED-DF field equations,  
$\mathrm{C}_{\alpha}{}^{\beta}\!= G_{\alpha}{}^{\beta}\!- 8\pi(T_{\alpha}{}^{\beta})\!_{_{_{ E \! D}}}\!- 8\pi(T_{\alpha}{}^{\beta})\!_{_{_{D\!F}}} \! =0$, in several cases of interest.  


%
Therefore, the GR-ED-DF equations in the Lagrangian formalism [GR-$\mathcal{L}(\mathcal{F})$-DF]  
for a general static spherically symmetric spacetime \eqref{SSSmet} are 
\begin{eqnarray}
&&\mathrm{C}_{t}{}^{t}  = 0 \quad\Rightarrow\quad -\frac{b'}{r^{2}}
 + \frac{2q^{2}_{_{\mathcal{E}}} }{ r^{4}\mathcal{L}_{\mathcal{F}}} + 2\mathcal{L} + 8\pi\rho\!_{_{_{DF}}} = 0,
\label{LFeqt}\\ 
&&\mathrm{C}_{r}{}^{r}  = 0 \quad\Rightarrow\quad \frac{2}{r}\!\!\left(1-\frac{b}{r}\right)\!\!\Phi' - \frac{b}{r^{3}}
+ \frac{ 2q^{2}_{_{\mathcal{E}}}  }{r^{4}\mathcal{L}_{\mathcal{F}}} + 2\mathcal{L} = 0,
\label{LFeqr}   
\end{eqnarray}
and the components $\mathrm{C}_{a}{}^{a}$, with $a= \{\theta, ~\varphi\}$, both reduce to   
\begin{equation}\label{LFeqphi}
\mathrm{C}_{a}{}^{a} = 0 
\quad\Rightarrow\quad
\left(1- \frac{b}{r}\right)\!\!\left( \Phi'^{2} + \Phi''\right) + \frac{ 2r - b - rb' }{ 2r^{2} }\Phi' - \frac{b'}{ 2r^{2} } + \frac{b}{ 2r^{3} } -\frac{2q^{2}_{_{\mathcal{B}}}}{r^4}\mathcal{L}_{\mathcal{F}} + 2\mathcal{L} = 0.  
\end{equation}      
Whereas the GR-ED-DF equations in the Hamiltonian formalism [GR-$\mathcal{H}(\mathcal{P})$-DF] for a spacetime metric \eqref{SSSmet}, are
\begin{eqnarray}
&&\mathrm{C}_{t}{}^{t}  = 0 \quad\Rightarrow\quad -\frac{b'}{r^{2}}
+ \frac{ 2q^{2}_{_{\mathcal{B}}} }{r^{4}\mathcal{H}_{\mathcal{P}}} - 2\mathcal{H} + 8\pi\rho\!_{_{_{DF}}} = 0,
\label{HPeqt}\\    
&&\mathrm{C}_{r}{}^{r}  = 0 \quad\Rightarrow\quad \frac{2}{r}\!\!\left(1-\frac{b}{r}\right)\!\!\Phi' - \frac{b}{r^{3}}
 + \frac{ 2q^{2}_{_{\mathcal{B}}}  }{r^{4}\mathcal{H}_{\mathcal{P}}} - 2\mathcal{H} = 0,
\label{HPeqr} %
\end{eqnarray}
and the components $\mathrm{C}_{a}{}^{a}$ with $a= \{\theta, ~\varphi\}$, lead to 
\begin{equation}\label{HPeqphi}
\mathrm{C}_{a}{}^{a} = 0 
\quad\Rightarrow\quad
\left(1- \frac{b}{r}\right)\!\!\left( \Phi'^{2} + \Phi''\right) + \frac{ 2r - b - rb' }{ 2r^{2} }\Phi' - \frac{b'}{ 2r^{2} } + \frac{b}{ 2r^{3} } -\frac{2q^{2}_{_{\mathcal{E}}}}{r^4}\mathcal{H}_{\mathcal{P}} - 2\mathcal{H} = 0. 
\end{equation} 

\subsubsection{ $\mathcal{FP}$ duality transformations in GR-ED-DF}  
We now consider two arbitrary (linear or nonlinear) electrodynamics theories, one
determinate by the electromagnetic Lagrangian density $\mathcal{L}^{^{(I)}}=\mathcal{L}^{^{(I)}}\!\!(\mathcal{F})$, and the other by the electromagnetic Hamiltonian density $\mathcal{H}^{^{(I\!I)}}=\mathcal{H}^{^{(I\!I)}}\!\!(\mathcal{P})$, such that the system GR-$\mathcal{L}^{^{(I)}}\!\!(\mathcal{F})$-DF admits the following static and spherically symmetric solution 
\begin{equation}
\left\{g^{^{(I)}}_{\mu\nu}, ~q^{^{(I)}}_{_{\mathcal{B}}}, ~q^{^{(I)}}_{_{\mathcal{E}}}, ~\mathcal{L}^{^{(I)}}, ~\mathcal{L}^{^{(I)}}_{\mathcal{F}}, ~\rho^{^{(I)}}_{_{_{DF}}}\right\}  
\end{equation}
whereas the system GR-$\mathcal{H}^{^{(I\!I)}}\!\!(\mathcal{P})$-DF admits the static and spherically symmetric solution 
\begin{equation}
\left\{g^{^{(I\!I)}}_{\mu\nu}, ~q^{^{(I\!I)}}_{_{\mathcal{B}}}, ~q^{^{(I\!I)}}_{_{\mathcal{E}}}, ~\mathcal{H}^{^{(I\!I)}}, ~\mathcal{H}^{^{(I\!I)}}_{\mathcal{P}}, ~\rho^{^{(I\!I)}}_{_{_{DF}}}\right\}.
\end{equation}
Then by comparing Eqs.\!~\eqref{LFeqt}-\eqref{LFeqphi}
and \eqref{HPeqt}-\eqref{HPeqphi}, one can conclude that $g^{^{(I\!I)}}_{\mu\nu} = g^{^{(I)}}_{\mu\nu}$ provided that the following equalities are satisfied
\begin{equation}\label{FP_duality_tran}
\mathcal{H}^{^{(I\!I)}}\!\!(\mathcal{X}) = -\mathcal{L}^{^{(I)}}\!\!(-\mathcal{X}), \quad\quad \mathcal{H}^{^{(I\!I)}}_{\mathcal{P}}\!\!(\mathcal{X}) = \mathcal{L}^{^{(I)}}_{\mathcal{F}}\!\!(-\mathcal{X}), 
\end{equation}
and 
\begin{equation}\label{FP_duality_tranFuen} 
q^{^{(I\!I)}}_{_{\mathcal{B}}} = q^{^{(I)}}_{_{\mathcal{E}}}, \quad q^{^{(I\!I)}}_{_{\mathcal{E}}} = q^{^{(I)}}_{_{\mathcal{B}}}, \quad \rho^{^{(I\!I)}}_{_{_{DF}}}=\rho^{^{(I)}}_{_{_{DF}}}.
\end{equation}
{\it I.e.,} there is an electric-magnetic duality between the static and spherically symmetric solutions of GR-$\mathcal{L}^{^{(I)}}\!\!(\mathcal{F})$-DF with those of GR-$\mathcal{H}^{^{(I\!I)}}\!\!(\mathcal{P})$-DF. 
On the other hand, it is worth stressing that in the case when the Lagrangian density 
$\mathcal{L}^{^{(I)}}\!\!(\mathcal{F})$ and the Hamiltonian density $\mathcal{H}^{^{(I\!I)}}\!\!(\mathcal{P})$ associated by \eqref{FP_duality_tran}, are also related by
$\mathcal{H}^{^{(I\!I)}} = 2\mathcal{F}\mathcal{L}^{^{(I)}}_{\mathcal{F}} -  \mathcal{L}^{^{(I)}}$ with $\mathcal{P} = (\mathcal{L}_{\mathcal{F}})^{2}\mathcal{F}$ which implies that the (linear/nonlinear) electrodynamics theories $\mathcal{L}^{^{(I)}}\!\!(\mathcal{F})$ and $\mathcal{H}^{^{(I\!I)}}\!\!(\mathcal{P})$ are equivalent, yields that the so-called $\mathcal{FP}$ duality turns into the conventional electric-magnetic duality \cite{Salazar87,Salazar89} connecting solutions from the same theory; whereas 
in the opposite case, {\it i.e.,} $\mathcal{H}^{^{(I\!I)}} \neq 2\mathcal{F}\mathcal{L}^{^{(I)}}_{\mathcal{F}} -  \mathcal{L}^{^{(I)}}$ with $\mathcal{P} = (\mathcal{L}_{\mathcal{F}})^{2}\mathcal{F}$, the theories $\mathcal{L}^{^{(I)}}\!\!(\mathcal{F})$ and $\mathcal{H}^{^{(I\!I)}}\!\!(\mathcal{P})$ are nonequivalent, and then the $\mathcal{FP}$ duality connects solutions from different theories.

\section{Static and spherically symmetric spacetime solutions in GR-ED-DF}\label{secIII}
The main lines of the integration process yielding the solutions will be given in the following.\\
Using  the Eq. \eqref{LFeqt}, the function $b'(r)$ can be expressed as 
\begin{equation}\label{bpri}
b' = \frac{2q^{2}_{_{\mathcal{E}}}}{r^{2}\mathcal{L}_{\mathcal{F}}} + 2r^{2}\mathcal{L} + 8\pi r^{2}\rho_{\!\!_{_{_{D\!F}}}}.
\end{equation} 
While, from the Eq. \eqref{LFeqr}, for $b(r)$ we find
\begin{equation}\label{b_r} 
b = \frac{ 2r^{2}\Phi' + 2r^{3}\mathcal{L} \!+\!\frac{2q^{2}_{_{\mathcal{E}}} }{r\mathcal{L}_{\mathcal{F}} } 
 }{ 1 + 2r\Phi'}.    
\end{equation}
Now, substituting \eqref{bpri} and \eqref{b_r} into Eq. \eqref{LFeqphi}, and isolating $2\pi\rho_{\!\!_{_{_{D\!F}}}}$ one obtains
\begin{equation}\label{rho_gen} 
2\pi\rho_{\!\!_{_{_{D\!F}}}} \!=\! \frac{1}{\left(1\!+\!r\Phi'\right)\!\left(1\!+\!2r\Phi'\right)}\!\left\{  \mathcal{L} \!-\! \frac{q^{2}_{_{\mathcal{B}}} \mathcal{L}_{\mathcal{F}}}{r^{4}}  \!+\! \frac{1}{r}\!\left[\!1 \!-\! \frac{2 \mathcal{L}_{\mathcal{F}}}{r^{2}}\left( q^{2}_{_{\mathcal{B}}} \!+\! \frac{q^{2}_{_{\mathcal{E}}}}{\mathcal{L}^{2}_{\mathcal{F}}}\right)\!\right]\!\Phi' \!+\!  \frac{1}{2}\!\left(1 \!-\! 2r^{2}\mathcal{L} \!-\! \frac{2q^{2}_{_{\mathcal{E}}}}{r^{2}\mathcal{L}_{\mathcal{F}}} \right)\!\left(\Phi'' \!+\! 2\Phi'^{2}\right)
\right\}.
\end{equation}
On the other hand, using the electromagnetic invariant $\mathcal{F}$ given by \eqref{F_invariant}, and the fact that   
$\mathcal{L}' - \mathcal{F}'\mathcal{L}_{\mathcal{F}} = 0$ we arrive at the following equation 
\begin{equation}\label{LLFrelation}
q^{2}_{_{\mathcal{E}}}\left[ r (\mathcal{L}_{\mathcal{F}})' + 2\mathcal{L}_{\mathcal{F}} \right]
- \left(r^{5}\mathcal{L}' + 2q^{2}_{_{\mathcal{B}}}\mathcal{L}_{\mathcal{F}}\right)\mathcal{L}_{\mathcal{F}}^{2} = 0.    
\end{equation}
The Eqs. \eqref{bpri} and \eqref{b_r} are independent. 
In fact, defining $\Psi$ as the difference between $b'(r)$ [of the Eq. \eqref{bpri}] and the derivative of 
$b(r)$ [given by Eq \eqref{b_r}] with respect to $r$ [using \eqref{LLFrelation} to express $(\mathcal{L}_{\mathcal{F}})'$ in terms of  $\mathcal{L}'$ and $\mathcal{L}_{\mathcal{F}}$], yields
\begin{equation}
\Psi\!=\!\frac{4r^{3}\Phi'}{\left(1\!+\!r\Phi'\right)\!\left(1\!+\!2r\Phi'\right)^{2}}\!\left\{\!\mathcal{L} \!-\! \frac{q^{2}_{_{\mathcal{B}}}\mathcal{L}_{\mathcal{F}}}{r^{4}} \!+\! \frac{1}{r}\!\left[\! 1 \!-\! \frac{2\mathcal{L}_{\mathcal{F}} }{r^{2}}\left(\!q^{2}_{_{\mathcal{B}}} \!+\! \frac{q^{2}_{_{\mathcal{E}}} }{\mathcal{L}^{2}_{\mathcal{F}}}\!\right)\!\right]\!\Phi' \!+\! \frac{1}{2}\!\left(\! 1  \!-\! 2r^{2}\mathcal{L} \!-\! \frac{2q^{2}_{_{\mathcal{E}}}}{r^{2}\mathcal{L}_{\mathcal{F}}}\!\right)\!\!\left(\!\Phi'' \!+\! 2\Phi'^{2}\right)\!\right\}  
\end{equation}
which, using Eq. \eqref{rho_gen}, can be written as $\Psi\!=\!\frac{8\pi r^{3}}{1 + 2r\Phi'\!(r)} \Phi'\!(r)\rho_{\!\!_{_{_{D\!F}}}}\!(r)$.
Thus, for Eqs. \eqref{bpri} and \eqref{b_r} to be equivalent, 
it is necessary that $\Psi=0$, implying 
\begin{equation}\label{Phi_rho}
 \Phi'\!(r)~\rho_{\!\!_{_{_{D\!F}}}}\!\!(r) = 0 \quad\quad\quad\quad \forall r  
\end{equation}
which gives rise to two disjoint (nontrivial) branches. The branch with nontrivial   
redshift function $\left[\Phi(r)\neq constant \right]$:   
\begin{equation}\label{branch1}
\textup{(i)}\quad
\Phi'\!\neq\!0 \quad\textup{and}\quad \mathcal{L} \!-\! \frac{q^{2}_{_{\mathcal{B}}}\mathcal{L}_{\mathcal{F}}}{r^{4}} \!+\! \frac{1}{r}\!\left[\! 1 \!-\! \frac{2\mathcal{L}_{\mathcal{F}} }{r^{2}}\left(\!q^{2}_{_{\mathcal{B}}} \!+\! \frac{q^{2}_{_{\mathcal{E}}} }{\mathcal{L}^{2}_{\mathcal{F}}}\!\right)\!\right]\!\Phi' \!+\! \frac{1}{2}\!\left(\! 1 \!-\! 2r^{2}\mathcal{L} \!-\! \frac{2q^{2}_{_{\mathcal{E}}}}{r^{2}\mathcal{L}_{\mathcal{F}}}\!\right)\!\!\left(\!\Phi'' \!+\! 2\Phi'^{2}\right)\!= 0 \quad\!\!\Rightarrow\!\!\quad \rho_{\!\!_{_{_{D\!F}}}}\!=\!0    
\end{equation}
and the branch with trivial redshift function $\left[\Phi(r)=constant\right]$:
\begin{equation}\label{branch2}
\textup{(ii)}\quad
\Phi'\!=\!0 \quad\textup{and}\quad \mathcal{L} \!-\! \frac{q^{2}_{_{\mathcal{B}}}\mathcal{L}_{\mathcal{F}}}{r^{4}}\!\neq\!0 \quad\!\!\Rightarrow\!\!\quad  \rho_{\!\!_{_{_{D\!F}}}}\!=\!\frac{1}{2\pi}\left(\mathcal{L} \!-\! \frac{q^{2}_{_{\mathcal{B}}}\mathcal{L}_{\mathcal{F}}}{r^{4}}\right)\!\neq\!0.
\end{equation}
One can verify through direct calculation that the line element \eqref{SSSmet}, with $b(r)$ as defined in \eqref{b_r}, and a dust fluid with energy density in the form of \eqref{rho_gen} [both expressed in terms of
$\Phi(r)$, $\mathcal{L}(\mathcal{F})$ and their derivatives, using Eq. \eqref{LLFrelation} to represent
$(\mathcal{L}_{\mathcal{F}})'$ in terms of  $\mathcal{L}'$ and $\mathcal{L}_{\mathcal{F}}$ ], satisfies the system of equations \eqref{LFeqt}$-$\eqref{LFeqphi} solely for the scenarios described by Eqs. \eqref{branch1} and \eqref{branch2}.

{\bf Theorem I.} Static and spherically symmetric traversable wormhole solutions with nontrivial redshift functions are not allowed in the framework of GR-ED-DF gravity. 
\\
{\bf Proof.} For an arbitrary static and spherically symmetric spacetime with nontrivial redshift function in GR-ED-DF, the fulfillment of Eqs. \eqref{LFeqt}$-$\eqref{LFeqphi} and \eqref{branch1} is necessary.
For this case, using Eqs. \eqref{LFeqt} and \eqref{LFeqr}, we obtain  
\begin{equation}
 2\Phi'\!(r) = \left[\ln\left(1 - \frac{b(r)}{r}\right)\right]'   
\end{equation}
which may be integrated to yield the relation $e^{2\Phi(r)} = \left(1 - \frac{b(r)}{r}\right)e^{\gamma}$, being $\gamma$ an integration constant which without loss of generality can assume that $\gamma=0$. 
Therefore, for this case the metric \eqref{SSSmet} takes the form
\begin{equation}
\boldsymbol{ds}^{2} =  - \left(1 - \frac{b(r)}{r}\right)\boldsymbol{dt}^{2} + \left(1 - \frac{b(r)}{r}\right)^{-1}\boldsymbol{dr}^{2} + r^{2}\boldsymbol{d\Omega}^{2}.    
\end{equation}
This corresponds to the metric ansatz for a static and spherically symmetric nontraversable wormhole solution, as it possesses an event horizon at the throat $r_{0}=b(r_{0})$.

These types of configurations have been widely studied, giving rise to several black hole solutions. 
Some of them are: Reissner-Nordstrom black holes and interesting models of black holes free of curvature singularities (see, e.g., Refs. \cite{Ayon1998,Bronnikov_Ayon1998,Bronnikov2000} or Ref. \cite{canate2023} for a recent review). \\ 
This demonstrates that static and spherically symmetric T-WH solutions (with $\Phi(r)\neq constant$)  are not possible within the framework of GR-ED-DF. 

In the forthcoming section we will reveal that, within the framework of GR-ED-DF, the only permissible static and spherically symmetric T-WHs are exclusively the ultrastatic and spherically symmetric types, meaning those static and spherically symmetric T-WHs with a trivial redshift function ({\it i.e.,} with $\Phi(r)=constant$).

\section{Ultrastatic and spherically symmetric spacetime solutions in GR-ED-DF}\label{secIV}
In this section, for an arbitrary (linear or nonlinear) electrodynamics model described by 
a Lagrangian density $\mathcal{L}(\mathcal{F})$ [in the $(\mathcal{L},\mathcal{F})$-formalism], or by a Hamiltonian density $\mathcal{H}(\mathcal{P})$ [in the $(\mathcal{H},\mathcal{P})$-formalism], 
we present a simple procedure to derive the most general ultrastatic and spherically symmetric solution of GR-$\mathcal{L}(\mathcal{F})$-DF [or GR-$\mathcal{H}(\mathcal{P})$-DF].
The general solution is characterized by having electric $q_{_{\mathcal{E}}}$ and magnetic $q_{_{\mathcal{B}}}$ charges. However, due to the nonlinear electromagnetic field effect, the general solution does not necessarily connect with the particular cases: purely magnetic ($q_{_{\mathcal{E}}}=0\neq q_{_{\mathcal{B}}}$) and purely electric ($q_{_{\mathcal{E}}}\neq0=q_{_{\mathcal{B}}}$).
For that reason, the purely magnetic and purely electric cases will be obtained first. 

\subsection{ Purely magnetic solutions in the $(\boldsymbol{\mathcal{L}},\boldsymbol{\mathcal{F}})-$formalism } 

The relevant GR-$\mathcal{L}(\mathcal{F})$-DF field equations for the purely magnetic case are given by \eqref{LFeqt}$-$\eqref{LFeqphi} with $q_{_{\mathcal{E}}}=0$ and $q_{_{\mathcal{B}}}=q$. For this case, according to Eq.\!~\eqref{F_invariant}, the electromagnetic invariant $\mathcal{F}$ takes the form $\mathcal{F} = q^{2}/(2r^{4})$.  


Below we present a method that, starting from an arbitrary electromagnetic Lagrangian density $\mathcal{L}(\mathcal{F})$, generates a magnetically charged ultrastatic and spherically symmetric spacetime solution in GR-$\mathcal{L}(\mathcal{F})$-DF gravity. 
\\
{\it Method.} Starting from an arbitrary electromagnetic Lagrangian density $\mathcal{L}=\mathcal{L}(\mathcal{F})$, using \eqref{b_r}, \eqref{rho_gen}, and \eqref{branch2}, we can deduce that the metric 
\begin{equation}\label{UltraS_Solution} 
\boldsymbol{ds}^{2} =  - \boldsymbol{dt}^{2} + \frac{\boldsymbol{dr}^{2}}{ 1 - \frac{b(r)}{r} }  + r^{2}(\boldsymbol{d\theta}^{2}  + \sin^{2}\theta \boldsymbol{d\varphi}^{2}), \quad\quad 
\textup{with} \quad\quad b(r) = 2r^{3}\mathcal{L}, \quad\quad \mathcal{F} =  \frac{q^{2}}{2r^{4}},
\end{equation}
where  $q$ is a real parameters (representing the total magnetic charge of the spacetime $q_{_{\mathcal{B}}}=q$), and the dust fluid with rest energy density 
\begin{equation}\label{density_dust} 
2\pi \rho_{\!\!_{_{_{D\!F}}}} = \mathcal{L} - F_{\theta\varphi}F^{\theta\varphi}\mathcal{L}_{\mathcal{F}}  
= \mathcal{L} - \frac{q^{2}}{r^{4}}\mathcal{L}_{\mathcal{F}},
\end{equation} 
is a purely magnetic, ultrastatic and spherically symmetric spacetime 
solution of GR-$\mathcal{L}(\mathcal{F})$-DF theory. 
In other words, given an arbitrary electromagnetic Lagrangian $\mathcal{L}(\mathcal{F})$ 
[being $\mathcal{L}(\mathcal{F})$ a well-definite function of invariant $\mathcal{F}$]  
one can see that the metric \eqref{UltraS_Solution}, together with the dust energy density \eqref{density_dust}, satisfy the GR-$\mathcal{L}(\mathcal{F})$-DF field equations \eqref{LFeqt}$-$\eqref{LFeqphi} for the purely magnetic case [{\it i.e.,} electromagnetic field tensor \eqref{Fab_SSS} with \eqref{F_invariant} for $q_{_{\mathcal{E}}}=0$ and $q_{_{\mathcal{B}}}=q$].

{\it Applications:}

\subsubsection{Magnetically charged Ellis-Bronnikov wormhole in GR-Maxwell-DF gravity} 
To begin with, let us consider the Lagrangian density of Maxwell's electrodynamics 
\begin{equation}\label{LED_Mx}
\mathcal{L}_{_{_{\mathrm{LED}}}}\!(\mathcal{F}) = \mathcal{F}.    
\end{equation}
For this case, evaluating \eqref{UltraS_Solution}, yields the following 
purely magnetic Ellis-Bronnikov wormhole metric
\begin{equation}\label{EB_WH}
\boldsymbol{ds}^{2} =  - \boldsymbol{dt}^{2} + \left( 1 - \frac{q^{2}}{r^{2}}  \right)^{\!\!^{-1}}
\boldsymbol{dr}^{2} + r^{2}(\boldsymbol{d\theta}^{2}  + \sin^{2}\theta \boldsymbol{d\varphi}^{2}),
\end{equation}
whereas, according to Eq.\!~\eqref{density_dust}, the corresponding dust fluid for which this metric is an exact purely magnetic solution of the GR-$\mathcal{L}_{_{_{\mathrm{LED}}}}\!(\mathcal{F})$-DF field equations, has a negative energy density given by 
\begin{equation}\label{dust_maxwell}
\rho_{\!\!_{_{_{D\!F}}}}\!\!(r) = -\frac{q^{2}}{ 4\pi r^{4}}.
\end{equation} 
This first example  connects with Ref. \cite{Shatskii08} in which the Ellis-Bronnikov wormhole metric \eqref{EB_WH}, originally derived in Refs. \cite{ellis73,Bronnikov73}, was reinterpreted in Ref. \cite{Shatskii08} as an exact solution of Einstein equations with a composite source
consisting of phantom dust [{\it i.e.,} dust fluid with negative energy density \eqref{dust_maxwell}]  
and a purely magnetic field obeying source-free Maxwell's equations\footnote{ I.e., the Eqs.\!~\eqref{e1} for the case of linear electrodynamics.}. 
Furthermore, recently in \cite{Bronnikov13}, it was shown that (in the context of general relativity) the source term of Ellis-Bronnikov wormhole also can be due to the contribution of a purely electric field obeying source-free Maxwell's equations and a dust fluid with negative energy density \eqref{dust_maxwell}. 

\subsubsection{ Magnetically charged wormhole in GR-EH-DF gravity}
In the following we shall consider the Euler-Heisenberg electrodynamics model in the approximation of the weak-field limit 
\begin{equation}\label{EH}
\mathcal{L}_{_{_{\mathrm{EH}}}}\!(\mathcal{F})
= \mathcal{F} - \frac{\mu^{2}}{2}\mathcal{F}^{2}.  
\end{equation}
Notice that the electrodynamics model \eqref{EH} satisfies the correspondence to Maxwell theory ({\it 
i.e.,} $\mathcal{L}_{_{_{\mathrm{EH}}}} \approx \mathcal{F}$ as $\mathcal{F}\approx0$).

Evaluating  \eqref{UltraS_Solution} from equation \eqref{EH}, one gets the following spacetime metric   
\begin{equation}\label{TWH_EH} 
\boldsymbol{ds}^{2} =  - \boldsymbol{dt}^{2} + \left( 1 - \frac{q^{2}}{r^{2}}  
+ \frac{ \mu^{2} q^{4}  
}{4r^{6}} \right)^{\!\!^{-1}} \boldsymbol{dr}^{2} + r^{2}(\boldsymbol{d\theta}^{2}  + \sin^{2}\theta \boldsymbol{d\varphi}^{2}),
\end{equation}
while, according to \eqref{density_dust},  the corresponding phantom\footnote{This means a dust fluid with negative energy density.} dust energy density  
for which this metric is an exact purely magnetic solution of the GR-$\mathcal{L}_{_{_{\mathrm{EH}}}}\!(\mathcal{F})$-DF field equations 
is given by
\begin{equation}\label{dust_EH} 
\rho_{\!\!_{_{_{D\!F}}}}\!\!(r) = -\frac{q^{2}}{4 \pi r^{4}}\!\!\left(1  - \frac{3 \mu^{2} q^{2}}{4r^{4}}\right)
\end{equation}  
The ultrastatic spacetime metric \eqref{TWH_EH}    
admits a T-WH interpretation since satisfies the properties \eqref{TWC1}$-$\eqref{AFWH} 
with wormhole throat radius $r_{_{0}}=(\mu^{2}q^{4}/4)^{^{\frac{1}{6}}} x_{_{0}}$,  
with $s =  (4q^{2}/\mu^{2})^{^{\frac{1}{3}}} > 3/(2)^{2/3}$ and $x_{_{0}}$ a positive number such that  
\begin{equation}
x^{2}_{_{0}} = \frac{s}{3} + \frac{1}{6}\left[-108 + 8s^{3} + 12\left(-12s^{3} + 81\right)^{\!\frac{1}{2}}\right]^{\!\frac{1}{3}} + \frac{2}{3}s^{2}\!\left[ -108 + 8s^{3} + 12\left(-12s^{3} + 81\right)^{\!\frac{1}{2}}\right]^{\!-\frac{1}{3}}.   
\end{equation}
It is evident that in the asymptotic region of the weak electromagnetic field ({\it i.e.,} $\mathcal{F}\approx0$ equivalent to $q\approx0$ or $r\sim\infty$), the GR-EH-DF T-WH solution [Eqs.\!~\eqref{EH}$-$\eqref{dust_EH}] reduces to the GR-Maxwell-DF T-WH solution [Eqs.\!~\eqref{LED_Mx}$-$\eqref{dust_maxwell}]. 

\subsubsection{ Magnetically charged wormhole in GR-BI-DF gravity }

Let us now consider the Born-Infeld electrodynamics model
\begin{equation}\label{BI}
\mathcal{L}_{_{\mathrm{BI}}}\!(\mathcal{F})
= 4\beta^{2} \left( -1 + \sqrt{ 1 + \frac{\mathcal{F}}{2\beta^{2}} }~\right),
\end{equation}
where  $\beta$ is a constant that has the physical interpretation of a critical field
strength \cite{BI}. \\
For this case, the metric Eqs.\!~\eqref{UltraS_Solution} takes the form 
\begin{equation}\label{TWH_BI}
\boldsymbol{ds}^{2} =  - \boldsymbol{dt}^{2} + \left[ 1  + 8\beta^{2} \left( r^{2}- \sqrt{ r^{4} + \frac{q^{2}}{4\beta^{2}} } ~ \right) \right]^{\!\!^{-1}} \boldsymbol{dr}^{2}
+ r^{2}(\boldsymbol{d\theta}^{2}  + \sin^{2}\theta \boldsymbol{d\varphi}^{2}).
\end{equation}  
According to \eqref{density_dust}, the corresponding dust fluid for which this metric is an exact purely magnetic solution of the GR-$\mathcal{L}_{_{\mathrm{BI}}}\!(\mathcal{F})$-DF field equations, has a negative energy density given by
\begin{equation}\label{dust_BI}
\rho_{\!\!_{_{_{D\!F}}}}\!\!(r) = \frac{2\beta^{2}}{\pi} \left( \frac{1}{\sqrt{ 1 + \frac{q^{2}}{4\beta^{2}r^{4}} } } - 1 \right).
\end{equation}
The metric \eqref{TWH_BI}, as long as $q^{2} > \frac{1}{16\beta^{2}}$, admits a T-WH interpretation since satisfies the properties \eqref{TWC1}$-$\eqref{AFWH} 
with wormhole throat radius 
\begin{equation}
r_{_{0}}= \sqrt{ q^{2} - \frac{1}{16\beta^{2}}  }.
\end{equation}
As in the previous case, it is evident that in the asymptotic region of the weak electromagnetic field ({\it i.e.,} $\mathcal{F}\approx0$ equivalent to $q\approx0$ or $r\sim\infty$), the GR-BI-DF T-WH solution [Eqs.\!~\eqref{BI}$-$\eqref{dust_BI}] reduces to the GR-Maxwell-DF T-WH solution [Eqs.\!~\eqref{LED_Mx}$-$\eqref{dust_maxwell}]. 

It is noteworthy that the electromagnetic sources of purely magnetic T-WHs\footnote{These are the purely magnetic GR-Maxwell-DF T-WH (or purely magnetic Ellis-Bronnikov wormhole) given by Eq.\!~\eqref{EB_WH}; the purely magnetic GR-EH-DF T-WH given by Eq.\!~\eqref{TWH_EH}; and the purely magnetic GR-BI-DF T-WH given by Eq.\!~\eqref{TWH_BI}.} derived in this subsection satisfy the NEC and WEC\footnote{The electromagnetic sources given by Eqs.\!~\eqref{LED_Mx}, \eqref{EH} and \eqref{BI}, corresponding to the purely magnetic T-WH solutions \eqref{EB_WH}, \eqref{TWH_EH} and \eqref{TWH_BI} respectively, satisfy the conditions \eqref{NEC_NLED} and \eqref{WEC_NLED}.}. On the other hand, because dust fluids with negative rest energy densities\footnote{These are \eqref{dust_maxwell}, \eqref{dust_EH} and \eqref{dust_BI}.} also make up the source of these T-WHs,
it is concluded that phantom dust fluids are solely responsible for the NEC and WEC violations\footnote{Fluids having negative rest energy density obviously contradict the requirement given by Eq.\!~\eqref{NEC_SF}.}
that are required for wormhole traversability.

Before finishing this subsection, it should be noted that, recently in Ref. \cite{Canate2019} the  Ellis-Bronnikov wormhole metric \eqref{EB_WH} 
was reinterpreted as a purely magnetic nonexotic T-WH solution of Einstein-scalar-Gauss-Bonnet coupled to Maxwell electrodynamics. 
Subsequently in Ref. \cite{Pcanate2023} the metrics \eqref{TWH_EH} and \eqref{TWH_BI} were introduced
as purely magnetic nonexotic T-WH solutions of Einstein-Gauss-Bonnet (with variable Gauss-Bonnet coefficient) coupled to Euler-Heisenberg and Born-Infeld nonlinear electrodynamics, respectively. 

On the other hand, with exception of \cite{Bronnikov13}, traversable wormhole
spacetimes (of the Morris-Thorne type) with electric charge are scarce. Below, in the context of general relativity, several electrically charged traversable wormhole solutions in GR-ED-DF gravity are obtained.

\subsection{Purely electric solutions in the $(\boldsymbol{\mathcal{H}},\boldsymbol{\mathcal{P}})-$formalism}

The relevant GR-$\mathcal{H}(\mathcal{P})$-DF equations for the purely electric case are given by Eqs.\!~\eqref{HPeqt}$-$\eqref{HPeqphi} with $q_{_{\mathcal{B}}}=0$ and $q_{_{\mathcal{E}}}=q$, where according to Eq.\!~\eqref{P_invariant} for this case the invariant $\mathcal{P}$ takes the form $\mathcal{P} = -q^{2}/(2r^{4})$.

Below, by applying the $\mathcal{FP}$ duality transformation Eqs.\!~\eqref{FP_duality_tran}-\eqref{FP_duality_tranFuen}, to the
purely magnetic GR-$\mathcal{L}(\mathcal{F})$-DF solution \eqref{UltraS_Solution}$-$\eqref{density_dust}, a method to generate ultrastatic, spherically symmetric and purely electric GR-$\mathcal{H}(\mathcal{P})$-DF solution
[with arbitrary $\mathcal{H(P)}$] is obtained.

{\it Method.} Starting from an arbitrary electromagnetic Hamiltonian density  
$\mathcal{H}=\mathcal{H}(\mathcal{P})$, the following metric
\begin{equation}\label{HPUltraS_Solution} 
\boldsymbol{ds}^{2} =  - \boldsymbol{dt}^{2} + \frac{\boldsymbol{dr}^{2}}{ 1 - \frac{b(r)}{r} }  + r^{2}(\boldsymbol{d\theta}^{2}  + \sin^{2}\theta \boldsymbol{d\varphi}^{2}), \quad\quad 
\textup{with} \quad\quad b(r) = -2r^{3}\mathcal{H}, 
\quad\quad \mathcal{P} =  -\frac{q^{2}}{2r^{4}},
\end{equation}
where  $q$ is a real parameters (representing the total electric charge of the spacetime $q_{_{\mathcal{E}}}=q$), and a dust fluid with rest energy density
\begin{equation}\label{HPdensity_dust} 
2\pi \rho_{\!\!_{_{_{D\!F}}}} = P_{tr}P^{tr}\mathcal{H}_{\mathcal{P}} - \mathcal{H} = - \frac{q^{2}}{r^{4}}\mathcal{H}_{\mathcal{P}} - \mathcal{H}. 
\end{equation}
is a purely electric, ultrastatic and spherically symmetric spacetime solution of GR-$\mathcal{H}(\mathcal{P})$-DF theory.\\
In similarity to the purely magnetic case Eqs.\!~\eqref{UltraS_Solution}$-$\eqref{density_dust}, given an arbitrary $\mathcal{H}(\mathcal{P})$ [being $\mathcal{H}(\mathcal{P})$ a well-definite function of invariant $\mathcal{P}$] one can see that the metric \eqref{HPUltraS_Solution} together with the dust energy density \eqref{HPdensity_dust} and electromagnetic field \eqref{P_invariant}, satisfy the GR-$\mathcal{H}(\mathcal{P})$-DF equations \eqref{HPeqt}$-$\eqref{HPeqphi} for the purely electric case [ i.e.,  nonzero components of $P_{\alpha\beta}$ given by \eqref{P_invariant} with $q_{_{\mathcal{E}}}=q$ and $q_{_{\mathcal{B}}}=0$].

{\it Applications: }
\subsubsection{Electrically charged Ellis-Bronnikov wormhole in GR-Maxwell-DF gravity} 

Using the $\mathcal{FP}$ duality transformation \eqref{FP_duality_tran} in 
$\mathcal{L}_{_{_{\mathrm{LED}}}}= \mathcal{F}$,  
one arrives at the following structural function: 
\begin{equation}\label{LED_Mx_HP}
\mathcal{H}_{_{_{\mathrm{LED}}}}\!(\mathcal{P}) = \mathcal{P}.    
\end{equation}
For this structural function, evaluating  Eq.\!~\eqref{HPUltraS_Solution}, 
yields the following metric
\begin{equation}\label{EB_maxwell_HP}
\boldsymbol{ds}^{2} =  - \boldsymbol{dt}^{2} + \left( 1 - \frac{q^{2}}{r^{2}}  \right)^{\!\!^{-1}}
\boldsymbol{dr}^{2} + r^{2}(\boldsymbol{d\theta}^{2}  + \sin^{2}\theta \boldsymbol{d\varphi}^{2}),
\end{equation}
whereas, according to Eq.\!~\eqref{HPdensity_dust}, the corresponding phantom dust energy density for which this metric is an exact purely electric traversable wormhole solution of the GR-$\mathcal{H}_{_{_{\mathrm{LED}}}}\!(\mathcal{P})$-DF gravity in the $(\mathcal{H},\mathcal{P})-$formalism [i.e., the Eqs.\!~\eqref{HPeqt}$-$\eqref{HPeqphi} with $\mathcal{H}(\mathcal{P})$ given by \eqref{LED_Mx_HP}] 
is 
\begin{equation}\label{dust_maxwell_HP}
\rho_{\!\!_{_{_{D\!F}}}}\!(r) = -\frac{q^{2}}{ 4\pi r^{4}}.
\end{equation}
It is important to  
emphasize that, even though \eqref{EB_WH} and \eqref{EB_maxwell_HP}
are the same metric,
the solutions \eqref{LED_Mx}-\eqref{dust_maxwell} and \eqref{LED_Mx_HP}-\eqref{dust_maxwell_HP} are different.  
This is because \eqref{LED_Mx}-\eqref{dust_maxwell} is a purely magnetic solution [i.e., with $q_{_{\mathcal{E}}}=0$ and $q_{_{\mathcal{B}}}=q$]  
whereas \eqref{LED_Mx_HP}$-$\eqref{dust_maxwell_HP} is a purely electric solution [i.e., with $q_{_{\mathcal{E}}}=q$ and $q_{_{\mathcal{B}}}=0$].  
Nevertheless, the theories $\mathcal{L}_{_{_{\mathrm{LED}}}}\!(\mathcal{F})$ and $\mathcal{H}_{_{_{\mathrm{LED}}}}\!(\mathcal{P})$ are equivalent because they are related by the Legendre transformation \eqref{actionHP}. 
Therefore, in this case the $\mathcal{F}\mathcal{P}$ duality connects different 
solutions from the same theory and becomes the conventional electric-magnetic duality \cite{Salazar87,Salazar89}.

\subsubsection{ Electrically charged wormhole in GR-NLED-DF gravity }

Applying \eqref{FP_duality_tran} in \eqref{EH} yields the following structural function 
\begin{equation}\label{EH_HP}
\mathcal{H}(\mathcal{P})  
= \mathcal{P} + \frac{ \mu^{2} }{2}\mathcal{P}^{2}   
\end{equation}
For this case, evaluating Eq.\!~\eqref{HPUltraS_Solution}, we find  
the following ultrastatic metric   
\begin{equation}\label{TWH_EH_HP} 
\boldsymbol{ds}^{2} =  - \boldsymbol{dt}^{2} + \left( 1 - \frac{q^{2}}{r^{2}} + \frac{ \mu^{2} q^{4}  
}{4r^{6}} \right)^{\!\!^{-1}} \boldsymbol{dr}^{2} + r^{2}(\boldsymbol{d\theta}^{2}  + \sin^{2}\theta \boldsymbol{d\varphi}^{2}),
\end{equation}
while, according to \eqref{HPdensity_dust},  the corresponding phantom dust energy density 
for which this metric is an exact purely electric traversable wormhole solution of the GR-$\mathcal{H}(\mathcal{P})$-DF field equations is given by
\begin{equation}\label{dust_EH_HP}
\rho\!_{\!\!_{_{_{D\!F}}}}\!(r) = -\frac{q^{2}}{4\pi r^{4}}\!\!\left(1 - \frac{3 \mu^{2} q^{2}}{4r^{4}}\right).
\end{equation}
Here, in contrast to the above case (Maxwell's electrodynamics), the Lagrangian \eqref{EH} and the Hamiltonian \eqref{EH_HP} are not related by \eqref{actionHP}$-$\eqref{def_P} and hence they describes different nonlinear electrodynamics theories. Therefore, in this case the $\mathcal{FP}$ duality connects  
solutions [which share the same metric \eqref{TWH_EH}$\equiv$\eqref{TWH_EH_HP}] of different GR-ED-DF systems characterized by two different nonlinear electrodynamics theories given by \eqref{EH} and \eqref{EH_HP}. 
 
\subsubsection{Electrically charged wormhole in GR-BI-DF gravity}

Applying the transformation \eqref{FP_duality_tran} in the Born-Infeld electrodynamics model \eqref{BI}, one arrive at the structural function
\begin{equation}\label{BI_HP}
\mathcal{H}_{_{\mathrm{BI}}}\!(\mathcal{P})
= 4\beta^{2} \left( 1 - \sqrt{ 1 - \frac{\mathcal{P}}{2\beta^{2}} }~\right).
\end{equation}
For this case, evaluating Eqs.\!~\eqref{HPUltraS_Solution}, yields the following
ultrastatic metric 
\begin{equation}\label{TWH_BI_HP}
\boldsymbol{ds}^{2} =  - \boldsymbol{dt}^{2} + \left[ 1  + 8\beta^{2} \left( r^{2}- \sqrt{ r^{4} + \frac{q^{2}}{4\beta^{2}} } ~ \right) \right]^{\!\!^{-1}} \boldsymbol{dr}^{2}
+ r^{2}(\boldsymbol{d\theta}^{2}  + \sin^{2}\theta \boldsymbol{d\varphi}^{2}).
\end{equation}
Whereas, according to \eqref{HPdensity_dust}, the corresponding phantom dust energy density  
for which this metric is an exact purely electric traversable wormhole solution of the GR-$\mathcal{H}_{_{\mathrm{BI}}}(\mathcal{P})$-DF  field equations 
is given by
\begin{equation}\label{dust_BI_HP}
\rho\!_{\!\!_{_{_{D\!F}}}}\!(r) = \frac{2\beta^{2}}{\pi}\!\!\left( \frac{1}{\sqrt{ 1 + \frac{q^{2}}{4\beta^{2}r^{4}} } } - 1 \right).
\end{equation}  %
As in the GR-Maxwell-DF case, \eqref{TWH_BI} and \eqref{TWH_BI_HP} are the same metric, however 
\eqref{BI}$-$\eqref{dust_BI} and \eqref{BI_HP}$-$\eqref{dust_BI_HP} are not the same solution:  
\eqref{BI}$-$\eqref{dust_BI} is a purely magnetic solution, whereas \eqref{BI_HP}$-$\eqref{dust_BI_HP} is a purely electric solution. Nevertheless, the theories $\mathcal{L}_{_{_{\mathrm{BI}}}}\!(\mathcal{F})$ and $\mathcal{H}_{_{_{\mathrm{BI}}}}\!(\mathcal{P})$ are equivalent because they are related by the Legendre transformation \eqref{actionHP}. That is to say, $\mathcal{L}_{_{_{\mathrm{BI}}}}\!(\mathcal{F})$ and $\mathcal{H}_{_{_{\mathrm{BI}}}}\!(\mathcal{P})$ admit the same solutions. Indeed, \eqref{TWH_BI_HP}$-$\eqref{dust_BI_HP} is also a purely electric solution of GR-$\mathcal{L}_{_{_{\mathrm{BI}}}}\!(\mathcal{F})$-DF, and \eqref{TWH_BI}$-$\eqref{dust_BI} is also a purely magnetic solution of GR-$\mathcal{H}_{_{_{\mathrm{BI}}}}\!(\mathcal{P})$-DF. 

It is noteworthy that the electromagnetic sources of purely electric T-WHs\footnote{These are the purely electric GR-Maxwell-DF T-WH (or purely electric Ellis-Bronnikov wormhole) given by Eq.\!~\eqref{EB_maxwell_HP}; the purely electric GR-NLED-DF T-WH given by Eq.\!~\eqref{TWH_EH_HP}; and the purely electric GR-BI-DF T-WH given by Eq.\!~\eqref{TWH_BI_HP}.} derived in this subsection satisfy the NEC and WEC\footnote{The electromagnetic sources given by Eqs.\!~\eqref{LED_Mx_HP}, \eqref{EH_HP} and \eqref{BI_HP}, corresponding to the purely magnetic T-WH solutions \eqref{EB_maxwell_HP}, \eqref{TWH_EH_HP} and \eqref{TWH_BI_HP} respectively, satisfy the conditions \eqref{NEC_NLEDHP} and \eqref{WEC_NLEDHP}.}. On the other hand, because dust fluids with negative rest energy densities\footnote{These are \eqref{dust_maxwell_HP}, \eqref{dust_EH_HP} and \eqref{dust_BI_HP}.} also make up the source of these T-WHs,
it is concluded that phantom dust fluids are solely responsible for the NEC and WEC violations 
that are required for wormhole traversability (for details, see Appendix).

In summary, all the ultrastatic and spherically symmetric traversable wormhole (of the Morris-Thorne type) solutions in GR-ED-DF obtained so far have been purely magnetic (i.e., $q_{_{\mathcal{E}}}=0\neq q_{_{\mathcal{B}}}$) or purely electric (i.e., $q_{_{\mathcal{E}}}\neq0=q_{_{\mathcal{B}}}$). Below, for the first time, Morris-Thorne traversable wormholes with both electric and magnetic charge (in the context of GR-ED-DF) will be presented.


\subsection{Solutions with electric and magnetic charges in the $(\boldsymbol{\mathcal{L}},\boldsymbol{\mathcal{F}})$ and  $(\boldsymbol{\mathcal{H}},\boldsymbol{\mathcal{P}})$ formalisms}
In this subsection, we shall derive the most general ultrastatic and spherically symmetric spacetime solution of GR-ED-DF equations.
In order to this, let us now consider the GR-$\mathcal{L}(\mathcal{F})$-DF equations \eqref{LFeqt}$-$\eqref{LFeqphi} for the most general case in which both charges ($q_{_{\mathcal{E}}}$ and $q_{_{\mathcal{B}}}$) are nonzero. For this type of configuration, the difficulty in finding the solution is that now according to Eq.\!~\eqref{F_invariant} or \eqref{P_invariant}, the electromagnetic invariant $\mathcal{F}$ or $\mathcal{P}$, is not always an explicitly function of $r$. 

{\it Method in the $(\mathcal{L},\mathcal{F})$-formalism.} Starting from an arbitrary $\mathcal{L}=\mathcal{L}(\mathcal{F})$ model, using \eqref{b_r}, \eqref{rho_gen} and \eqref{branch2}, we can deduce that the metric
\begin{equation}\label{UltraS_SolutionEM} 
\boldsymbol{ds}^{2} \!=\! - \boldsymbol{dt}^{2} \!+ \frac{\boldsymbol{dr}^{2}}{ 1 \!-\! \frac{b(r)}{r} }  \!+ r^{2}\boldsymbol{d\Omega}^{2},  \quad\quad b(r) \!=\! 2r^{3}\!\left( \mathcal{L} \!-\! F_{tr}F^{tr}\mathcal{L}_{\mathcal{F}} \right) \!=\! 2r^{3}\mathcal{L} \!+\! \frac{2q^{2}_{_{\mathcal{E}}}}{r\mathcal{L}_{\mathcal{F}}}, \quad\quad 
\mathcal{F}\!=\!\frac{1}{2r^{4}}\!\!\left(q^{2}_{_{\mathcal{B}}}  \!-\! \frac{ q^{2}_{_{\mathcal{E}}} }{ \mathcal{L}^{^{2}}_{\mathcal{F}} }\right), 
\end{equation} 
and a dust fluid with rest energy density
\begin{equation}\label{density_dust_EM_LF}  
2\pi \rho_{\!\!_{_{_{D\!F}}}} = \mathcal{L} - F_{\theta\varphi}F^{\theta\varphi}\mathcal{L}_{\mathcal{F}} 
=  \mathcal{L} - \frac{q^{2}_{_{\mathcal{B}}}}{r^{4}}\mathcal{L}_{\mathcal{F}},
\end{equation}
becomes the most general ultrastatic and spherically symmetric solution of the GR-$\mathcal{L}(\mathcal{F})$-DF equations \eqref{LFeqt}$-$\eqref{LFeqphi}. 
Whereas, regarding this construction in the $(\mathcal{H},\mathcal{P})$-formalism, 
applying the $\mathcal{FP}$-duality transformation to the generic ultrastatic and spherically symmetric solution \eqref{UltraS_SolutionEM}$-$\eqref{density_dust_EM_LF} of GR-$\mathcal{L(F)}$-DF equations, we obtain the following procedure.

{\it Method in the $(\mathcal{H},\mathcal{P})$-formalism.} Starting from an arbitrary $\mathcal{H}=\mathcal{H}(\mathcal{P})$ model, the following metric 
\begin{equation}\label{UltraS_SolutionEM_HP} 
\boldsymbol{ds}^{2} \!=\! - \boldsymbol{dt}^{2} \!+ \frac{\boldsymbol{dr}^{2}}{ 1 \!-\! \frac{b(r)}{r} }  \!+ r^{2}\boldsymbol{d\Omega}^{2},  
\quad\quad 
b(r) \!=\! 2r^{3}\!\left( -\mathcal{H} \!+\! P_{\theta\varphi}P^{\theta\varphi}\mathcal{H}_{\mathcal{P}} \right) \!=\! -2r^{3}\mathcal{H} \!+\! \frac{ 2q^{2}_{_{\mathcal{B}}} }{r\mathcal{H}_{\mathcal{P}}},  
\quad\quad
\mathcal{P} \!=\! \frac{1}{2r^{4}}\!\!\left(\frac{q^{2}_{_{\mathcal{B}}}}{\mathcal{H}^{^{2}}_{\mathcal{P}}}  \!-\! q^{2}_{_{\mathcal{E}}}\right),
\end{equation}
together with a dust fluid with rest energy density 
\begin{equation}\label{density_dust_EM_HP}   
2\pi \rho_{\!\!_{_{_{D\!F}}}} = P_{tr}P^{tr}\mathcal{H}_{\mathcal{P}} - \mathcal{H} = - \frac{q^{2}_{_{\mathcal{E}}}}{r^{4}}\mathcal{H}_{\mathcal{P}} - \mathcal{H}, 
\end{equation}
is the most general ultrastatic and spherically symmetric solution of GR-$\mathcal{H}(\mathcal{P})$-DF equations \eqref{HPeqt}$-$\eqref{HPeqphi}.

Summarizing, with this, we have obtained a new class of ultrastatic, spherically symmetric, electrically and magnetically charged spacetime solutions in GR-ED-DF gravity depending on an arbitrary Lagrangian $\mathcal{L}(\mathcal{F})$ in the $(\mathcal{L},\mathcal{F})$-formalism, or the an arbitrary Hamiltonian $\mathcal{H}(\mathcal{P})$ in the $(\mathcal{H},\mathcal{P})$-formalism. 

We present now an example with interesting particular cases.

\subsubsection{Novel traversable wormhole in GR-BI-DF gravity} 
 
Consider the Lagrangian density of Born-Infeld electrodynamics.  
For this case, evaluating \eqref{UltraS_SolutionEM}, one arrives at the metric 
\begin{equation}\label{WH_BI_EM}
\boldsymbol{ds}^{2} =  - \boldsymbol{dt}^{2} + \left\{ 1  + 8\beta^{2} \left[ r^{2}- \sqrt{ \left(r^{2} + \frac{q^{2}_{_{\mathcal{E}}}}{4\beta^{2}r^{2}}\right)\!\!\left(r^{2} + \frac{q^{2}_{_{\mathcal{B}}}}{4\beta^{2}r^{2}}\right) } ~ \right] \right\}^{\!\!^{-1}} \boldsymbol{dr}^{2}
+ r^{2}(\boldsymbol{d\theta}^{2}  + \sin^{2}\theta \boldsymbol{d\varphi}^{2}),
\end{equation}
together with the electromagnetic invariant 
\begin{equation}\label{BI_F_inv}
\mathcal{F} = \frac{ q^{2}_{_{\mathcal{B}}} - q^{2}_{_{\mathcal{E}}}  }{ 2 r^{4} + \frac{q^{2}_{_{\mathcal{E}}} }{2 \beta^{2}} }.   
\end{equation}
According to \eqref{density_dust_EM_LF}, the corresponding dust energy density for which this metric is the most general ultrastatic, spherically symmetric, electrically and magnetically charged
solution of GR-$\mathcal{L}_{_{_{\mathrm{BI}}}}\!(\mathcal{F})$-DF equations 
is given by  
\begin{equation}\label{dust_BI_EM} 
\rho_{\!\!_{_{_{D\!F}}}}\!\!(r) = \frac{2\beta^{2}}{\pi} \left( \frac{ 1 - \frac{q^{2}_{_{\mathcal{E}}}q^{2}_{_{\mathcal{B}}}}{16\beta^{4}r^{8}} }{\sqrt{ \left(1 + \frac{q^{2}_{_{\mathcal{E}}}}{4\beta^{2}r^{4}}\right)\!\!\left(1 + \frac{q^{2}_{_{\mathcal{B}}}}{4\beta^{2}r^{4}}\right) } } - 1 \right).
\end{equation}
It should be noted that for the following parameter setting, $q^{2}_{_{\mathcal{E}}}+ q^{2}_{_{\mathcal{B}}}\neq0\neq\beta$, the $g^{rr}$ metric component of the line element \eqref{WH_BI_EM} belongs to the class $\mathcal{C}^{\infty}$ for all $r\in(0,\infty)$, furthermore, it is such that $\lim\limits_{r\to\infty}g^{rr}(r) = 1$ and $\lim\limits_{r\to0} g^{rr}(r)\rightarrow -\infty$.
This behavior suggests the existence of a $r_{0}\in(0,\infty)$ such that $g^{rr}(r_{0})=0$. 
Consequently, within this parameter configuration, the metric \eqref{WH_BI_EM} describes the geometry of
wormhole with a throat radius $r_{0}$ and without event horizon.\\
An alternative way to obtain the solution \eqref{WH_BI_EM}$-$\eqref{dust_BI_EM},  
correspond to use the Born-Infeld electrodynamics in the $(\mathcal{H},\mathcal{P})-$formalism [i.e., GR-$\mathcal{H}(\mathcal{P})$-DF equations with the $\mathcal{H}(\mathcal{P})$ given by \eqref{BI_HP}]. 
Thus, evaluating the \eqref{UltraS_SolutionEM_HP}$-$\eqref{density_dust_EM_HP} for the Hamiltonian density \eqref{BI_HP}, one finds the metric \eqref{WH_BI_EM}, the invariant $\mathcal{P}$  
\begin{equation}\label{BI_P_inv}
\mathcal{P} = \frac{ q^{2}_{_{\mathcal{B}}} - q^{2}_{_{\mathcal{E}}}  }{ 2 r^{4} + \frac{q^{2}_{_{\mathcal{B}}} }{2 \beta^{2}} }.    
\end{equation}
and the dust energy density \eqref{dust_BI_EM}, respectively.  
\\
{\bf Energy conditions.} The BI electrodynamics source of the spacetime \eqref{WH_BI_EM} satisfies the NEC and WEC everywhere. A simple way to demonstrate this comes from considering the following two complementary cases:  
case $q^{2}_{_{\mathcal{B}}} \geq q^{2}_{_{\mathcal{E}}}$, and case $q^{2}_{_{\mathcal{B}}} \leq q^{2}_{_{\mathcal{E}}}$.  
For the case $q^{2}_{_{\mathcal{B}}} \geq q^{2}_{_{\mathcal{E}}}$,  
using conveniently the ($\mathcal{L},\mathcal{F}$)-formalism, according to \eqref{BI_F_inv} the electromagnetic invariant $\mathcal{F}$ is positive definite, and hence is easy to see that $\mathcal{L}_{_{\mathrm{BI}}}\!(\mathcal{F})$, Eq.\!~\eqref{BI}, satisfies \eqref{NEC_NLED} and \eqref{WEC_NLED}. While for the case $q^{2}_{_{\mathcal{B}}} \leq q^{2}_{_{\mathcal{E}}}$, 
employing the ($\mathcal{H},\mathcal{P}$)-formalism, according to \eqref{BI_P_inv} the invariant $\mathcal{P}$ is negative definite, and hence is easy to see that $\mathcal{H}_{_{\mathrm{BI}}}\!(\mathcal{P})$, Eq.\!~\eqref{BI_HP}, complies with \eqref{NEC_NLEDHP}  and \eqref{WEC_NLEDHP}.  \\
With regard to the dust fluid source of \eqref{WH_BI_EM}, according to Eq.\!~\eqref{dust_BI_EM}, has a negative rest energy density for all values of radial coordinates [disagree with Eq.\!~\eqref{NEC_SF}], signifying that this source violates the NEC and WEC everywhere in spacetime.\\
{\bf Particular cases.} Below, we will discuss the principal particular cases of metric \eqref{WH_BI_EM} 
to elucidate its geometrical significance. 
\begin{itemize}

\item{\bf Trivial case: Vanishing electromagnetic field}\\ 
For $q_{_{\mathcal{E}}}=0=q_{_{\mathcal{B}}}$ we have that $F_{tr}=0=F_{\theta\varphi}$ {\it(the electromagnetic field is turned off)}, whereas the dust energy density \eqref{dust_BI_EM} becomes zero and the line element \eqref{WH_BI_EM} reduces to the Minkowski metric. This is in accordance with the fact, according to  Eqs.\!~\eqref{UltraS_SolutionEM} or \eqref{UltraS_SolutionEM_HP}, the most general ultrastatic and spherically symmetric solution of a GR-ED-DF theory cannot have a nonzero  shape  
function without a linear (or nonlinear) electrodynamics source. 

\item{\bf Maxwell limit: Electrically and magnetically charged Ellis-Bronnikov WH}\\
For the Lagrangian density of Born-Infeld electrodynamics \eqref{BI} [or Hamiltonian density \eqref{BI_HP}], it is well known that the limit $\beta\rightarrow\infty$ guarantees the correspondence to the linear Maxwell theory, that is
\begin{equation}\label{LimBI}
\lim\limits_{\beta\to\infty}\mathcal{L}_{_{_{\mathrm{BI}}}} = \mathcal{L}_{_{_{\mathrm{LED}}}} = \mathcal{F} \quad\quad\textup{ or }\quad\quad\lim\limits_{\beta\to\infty}\mathcal{H}_{_{_{\mathrm{BI}}}} = \mathcal{H}_{_{_{\mathrm{LED}}}} = \mathcal{P}.    
\end{equation}
On the other hand, by a simple calculation, we find that when $\beta\to\infty$ 
the invariant \eqref{BI_F_inv} takes the form $\mathcal{F}=\frac{ q^{2}_{_{\mathcal{B}}} - q^{2}_{_{\mathcal{E}}} }{2r^{4}}$, whereas the expressions \eqref{WH_BI_EM} and \eqref{dust_BI_EM} respectively become,  
\begin{eqnarray}
\boldsymbol{ds}^{2} &=& - \boldsymbol{dt}^{2} + \left( 1  - \frac{ q^{2}_{_{\mathcal{E}}} + q^{2}_{_{\mathcal{B}}} }{r^{2}} \right)^{\!\!^{-1}} \boldsymbol{dr}^{2}
+ r^{2}(\boldsymbol{d\theta}^{2}  + \sin^{2}\theta \boldsymbol{d\varphi}^{2}),  \label{EB_WH_EM} \\
\rho_{\!\!_{_{_{D\!F}}}}\!\!(r) &=& - \frac{ q^{2}_{_{\mathcal{E}}} + q^{2}_{_{\mathcal{B}}} }{4\pi r^{4}}. \label{EB_WH_rho}   
\end{eqnarray} 
Since the metric \eqref{EB_WH_EM} has the structure of an Ellis-Bronnikov wormhole equipped with electric and magnetic charges, then it can be interpreted as an electrically and magnetically charged Ellis-Bronnikov wormhole with throat radius given by $r_{0} = \sqrt{q^{2}_{_{\mathcal{E}}} + q^{2}_{_{\mathcal{B}}}}$.
Alternatively, for consistent of the field equations, if one starts from
$\mathcal{L}_{_{_{\mathrm{LED}}}} = \mathcal{F}$ and computed  
 \eqref{UltraS_SolutionEM}$-$\eqref{density_dust_EM_LF} [or $\mathcal{H}_{_{_{\mathrm{LED}}}} = \mathcal{P}$ and computed \eqref{UltraS_SolutionEM_HP} and \eqref{density_dust_EM_HP}], yields  the same expressions \eqref{EB_WH_EM} and \eqref{EB_WH_rho}. 
Therefore, we conclude that the electrically and magnetically charged Ellis-Bronnikov wormhole spacetime \eqref{EB_WH_EM}, is the most general ultrastatic and spherically symmetric spacetime solution of general relativity coupled to Maxwell's electrodynamics and a dust fluid with negative energy density \eqref{EB_WH_rho}.
This generalizes \cite{Bronnikov13}, in which in the GR context
the source term $T_{\alpha\beta} = G_{\alpha\beta}/(8\pi)$
of an Ellis-Bronnikov wormhole as being due to two contributions 
$T_{\alpha\beta} = (T_{\alpha}{}^{\beta})\!_{_{_{M\!a\!x}}} + (T_{\alpha}{}^{\beta})\!_{_{_{ D\!F}}}$ where $(T_{\alpha}{}^{\beta})\!_{_{_{M\!a\!x}}}$ is attributable to a 
radial electric or magnetic field (but not both) and $(T_{\alpha}{}^{\beta})\!_{_{_{D\!F}}}$ is that of a dust fluid with negative energy density.

Summarizing, with this specific case, a new source of Ellis-Bronnikov wormhole is presented in the context of general relativity. This new source consists of a phantom dust fluid, and an electromagnetic field satisfying the source-free Eqs.\!~\eqref{e1} for the case of Maxwell electrodynamics. 
%
\item{\bf Ultrastatic and spherically symmetric GR-BI-DF solution 
with $\boldsymbol{q_{_{\mathcal{E}}}=0\neq q_{_{\mathcal{B}}}}$ }\\ 
For this case the purely magnetic solution  \eqref{BI}$-$\eqref{dust_BI} is recovered. 
\item{\bf Ultrastatic and spherically symmetric GR-BI-DF solution with $\boldsymbol{q_{_{\mathcal{E}}}\neq 0 = q_{_{\mathcal{B}}}}$}\\
For this case the purely electric solution  \eqref{BI_HP}$-$\eqref{dust_BI_HP} is recovered, but now written in the $(\mathcal{L},\mathcal{F})-$formalism. 
\item{\bf Ultrastatic and spherically symmetric GR-BI-DF solution with $\boldsymbol{q^{2}_{_{\mathcal{E}}}\neq q^{2}_{_{\mathcal{B}}}}$: New T-WH spacetime}

In this case, the shape function of the metric \eqref{WH_BI_EM}, 
\begin{equation}
b(r) = 8 \beta^{2}r \left[ \sqrt{ \left(r^{2} + \frac{q^{2}_{_{\mathcal{E}}}}{4\beta^{2}r^{2}}\right)\!\!\left(r^{2} + \frac{q^{2}_{_{\mathcal{B}}}}{4\beta^{2}r^{2}}\right) } - r^{2}  ~ \right]     
\end{equation}
such that the equation, $1 - \frac{b(r_{_{0}})}{r_{_{0}}}=0$, has a single solution for $r_{_{0}}>0$, which is 
\begin{equation}  
r_{_{0}} = \sqrt{ \frac{\left[(s^{2} + 1)v^{2}-1\right]^{2} + \left[   (s^{2} + 1)v^{2} + a - 1 \right]a }{48a\beta^{2}  }~ }~,   
\end{equation}    
where we have defined the auxiliary parameters 
 $s = |q_{_{\mathcal{B}}}|/|q_{_{\mathcal{E}}}|$, $v = 4\beta |q_{_{\mathcal{E}}}|$ and 
\begin{equation}
a\!=\!\left\{\!6 \sqrt{3\!\left[(\!s^{2}\!+\!1\!)^{^{3}}v^{6}\!-\!3(s^{4} \!-\! 7 s^{2} \!+\! 1)v^{4} \!+\! 3(s^{2}\!+\!1)v^{2}\!-\!1\right] }~\!s v^{2}\!+\! 
[(s^{2}\!+\!1)^{^{2}}v^{4}\!+\!3](s^{2}\!+\!1)v^{2}\!-\!3(s^{4}\!-\!16s^{2}\!+\!1)v^{4}  
\!-\!1\!\right\}^{\!\!\frac{1}{3}}\!.
\end{equation}
Evaluating 
$b'(r)$ at $r=r_{0}$, yields
\begin{equation}
b'(r)\Big|_{r=r_{_{0}}} = \frac{1 - 8 \beta^{2}r^{2}_{_{0}} - 8 q^{2}_{_{\mathcal{E}}}q^{2}_{_{\mathcal{B}}}r^{-4}_{_{0}} }{ 8\beta^{2}r^{2}_{_{0}} + 1},
\end{equation}
which satisfies, $b'(r)\Big|_{r=r_{_{0}}}<1$, for all $q_{_{\mathcal{E}}}$, $q_{_{\mathcal{B}}}$, $\beta$ such that  ($q^{2}_{_{\mathcal{E}}} + q^{2}_{_{\mathcal{B}}} \neq0\neq \beta$).   Hence, the metric \eqref{WH_BI_EM}, for $q^{2}_{_{\mathcal{E}}} + q^{2}_{_{\mathcal{B}}} \neq0\neq \beta$, describes a traversable wormhole geometry since it satisfies the conditions \eqref{TWC1}$-$\eqref{AFWH}.
%
%
\item {\bf Ultrastatic and spherically symmetric GR-BI-DF solution with
$\boldsymbol{q^{2}_{_{\mathcal{E}}} = q^{2}_{_{\mathcal{B}}}}$: Electrically and magnetically
charged Ellis-Bronnikov WH with vanishing electromagnetic invariant} 

Note that for the purely magnetic solution \eqref{UltraS_Solution}$-$\eqref{density_dust} the case $\mathcal{F}=0$ is only possible if $q_{_{\mathcal{B}}}=0$ [or for the purely electric solution \eqref{HPUltraS_Solution}$-$\eqref{HPdensity_dust} the $\mathcal{P}$=0 is only possible if $q_{_{\mathcal{E}}}=0$], 
implying the despaired of the wormhole structure, and the Minkowski spacetime geometry is recovered in both cases.
However, the present solution [i.e., the solution \eqref{WH_BI_EM}$-$\eqref{dust_BI_EM}] for the particular case  
$q^{2}_{_{\mathcal{E}}}=q^{2}_{_{\mathcal{B}}}\neq0$, 
is such that the line element \eqref{WH_BI_EM} reduces to the Ellis-Bronnikov wormhole metric
\begin{equation}\label{WH_BI_EB}
\boldsymbol{ds}^{2} =  - \boldsymbol{dt}^{2} + \left( 1  -   \frac{q^{2}_{_{\mathcal{E}}} + q^{2}_{_{\mathcal{B}}}}{r^{2}}  \right)^{\!\!^{-1}} \boldsymbol{dr}^{2}
+ r^{2}(\boldsymbol{d\theta}^{2}  + \sin^{2}\theta \boldsymbol{d\varphi}^{2}) \quad\quad\textup{with}\quad\quad q^{2}_{_{\mathcal{E}}} = q^{2}_{_{\mathcal{B}}}, 
\end{equation}
the electromagnetic invariant \eqref{BI_F_inv} becomes zero $\mathcal{F}\big|_{q^{2}_{_{\mathcal{E}}} = q^{2}_{_{\mathcal{B}}}} = ~ 0$, but the electric and magnetic fields be nonzero
[Eq. \eqref{Fab_SSS} with \eqref{F_invariant}], and the dust energy density \eqref{dust_BI_EM} takes the form
\begin{equation}\label{dust_BI_EB}
\rho_{\!\!_{_{_{D\!F}}}}\!\!(r)  = - \frac{q^{2}_{_{\mathcal{E}}} + q^{2}_{_{\mathcal{B}}}}{4\pi r^{4}} \quad\quad\textup{with}\quad\quad q^{2}_{_{\mathcal{E}}} = q^{2}_{_{\mathcal{B}}}.
\end{equation}
This mean that the most general ultrastatic and spherically symmetric solution of GR-BI-DF, Eqs.\!~\eqref{WH_BI_EM}$-$\eqref{dust_BI_EM}, for the particular case $q^{2}_{_{\mathcal{E}}} = q^{2}_{_{\mathcal{B}}}$ becomes an electrically and magnetically charged Ellis-Bronnikov wormhole with zero electromagnetic invariant. 

One can observe that the result [i.e., the metric \eqref{WH_BI_EB} and the phantom dust energy density \eqref{dust_BI_EB}] also can be obtained by restricting the most general ultastatic and spherically symmetric solution of GR-Maxwell-DF, Eqs.\!~\eqref{EB_WH_EM} and \eqref{EB_WH_rho}, to $q^{2}_{_{\mathcal{E}}} = q^{2}_{_{\mathcal{B}}}$. This suggests that several GR-ED-DF theories admit T-WH metrics of the electrically and magnetically charged Ellis-Bronnikov (with $\mathcal{F}=0$) type as particular solutions.

\end{itemize}

\subsection{Electrically and magnetically charged Ellis-Bronnikov wormholes (with zero electromagnetic invariant) in GR-ED-DF theories}

{\bf Theorem II.} For any GR-$\mathcal{L}(\mathcal{F})$-DF theory with $\mathcal{L}(\mathcal{F})$ such that
\begin{equation}\label{LLFcons}
\mathcal{L}\bigg|_{_{\mathcal{F}=0}} = ~ 0  \quad\quad \textup{and}\quad\quad \mathcal{L}_{_{\mathcal{F}}}\bigg|_{_{\mathcal{F}=0}} = ~ \sigma 
\end{equation}
where $\big|_{_{\mathcal{F}=0}}$ means restricted to $\mathcal{F}=0$, and $\sigma$ is a nonzero positive constant, the electrically and magnetically charged  
Ellis-Bronnikov wormhole metric  
\begin{equation}\label{sigma_WH_BI_EB}  
\boldsymbol{ds}^{2} =  - \boldsymbol{dt}^{2} + \left( 1  -   \frac{q^{2}_{_{\mathcal{E}}} + \sigma^{2}q^{2}_{_{\mathcal{B}}}}{\sigma r^{2}}  \right)^{\!\!^{-1}} \boldsymbol{dr}^{2}
+ r^{2}(\boldsymbol{d\theta}^{2}  + \sin^{2}\theta \boldsymbol{d\varphi}^{2}), 
\end{equation}
together with a dust fluid with negative energy density given by 
\begin{equation}\label{sigma_dust_BI_EB}    
\rho_{\!\!_{_{_{D\!F}}}}\!\!(r) = - \frac{q^{2}_{_{\mathcal{E}}} + \sigma^{2}q^{2}_{_{\mathcal{B}}}}{4\pi \sigma r^{4}} 
\end{equation}
for the particular case $q^{2}_{_{\mathcal{E}}} = \sigma^{2} q^{2}_{_{\mathcal{B}}}$,
is an exact solution of GR-$\mathcal{L}(\mathcal{F})$-DF equations \eqref{LFeqt}$-$\eqref{LFeqphi}, having
null electromagnetic invariant  
\begin{equation}
\mathcal{F}\bigg|_{q^{2}_{_{\mathcal{E}}} = \sigma^{2} q^{2}_{_{\mathcal{B}}}} =  ~ 0.
\end{equation}

{\bf Proof.} For the spacetime metric \eqref{sigma_WH_BI_EB},
the Einstein tensor has the following nonzero components: 
\begin{equation}\label{GREB_sigma}  
G_{t}{}^{t} = \frac{q^{2}_{_{\mathcal{E}}} + \sigma^{2}q^{2}_{_{\mathcal{B}}}}{\sigma r^{4}}, \quad\quad G_{r}{}^{r} = - \frac{q^{2}_{_{\mathcal{E}}} + \sigma^{2}q^{2}_{_{\mathcal{B}}}}{\sigma r^{4}}, \quad\quad G_{\theta}{}^{\theta} = G_{\varphi}{}^{\varphi}  = \frac{q^{2}_{_{\mathcal{E}}} + \sigma^{2}q^{2}_{_{\mathcal{B}}}}{\sigma r^{4}}.  
\end{equation}
Whereas, for an arbitrary static and spherically symmetric spacetime, the energy-momentum tensor of a dust fluid $(T_{\alpha}{}^{\beta})\!_{_{_{D\!F}}} = \rho_{\!\!_{_{_{D\!F}}}} u_{\alpha}u^{\beta}$ with rest energy density  $\rho_{\!\!_{_{_{D\!F}}}}$ given by 
\eqref{sigma_dust_BI_EB}, only has the following nonzero component:
\begin{equation}\label{DFEB_sigma}  
8\pi (T_{t}{}^{t})\!_{_{_{D\!F}}} = - 8\pi\rho\!_{_{_{DF}}} = \frac{2q^{2}_{_{\mathcal{E}}} + 2\sigma^{2}q^{2}_{_{\mathcal{B}}}}{ \sigma r^{4}}.  
\end{equation}
On the other hand, in the $(\mathcal{L},\mathcal{F})$-formalism for a Lagrangian density $\mathcal{L}(\mathcal{F})$ satisfying \eqref{LLFcons}, the nontrivial components of $(T_{\alpha}{}^{\beta})\!_{_{_{E\!D}}}$ for an arbitrary static, spherically symmetric, electrically and magnetically charged spacetime metric having null electromagnetic invariant $\mathcal{F}=0$, are 
\begin{equation}\label{LFEB_sigma}  
8\pi (T_{t}{}^{t})\!_{_{_{E\!D}}} = 8\pi (T_{r}{}^{r})\!_{_{_{E\!D}}}  = -\frac{2q^{2}_{_{\mathcal{E}}} }{\sigma r^{4}},  \quad\quad 8\pi (T_{\theta}{}^{\theta})\!_{_{_{E\!D}}} = 8\pi (T_{\varphi}{}^{\varphi})\!_{_{_{E\!D}}} =  \frac{2\sigma q^{2}_{_{\mathcal{B}}}}{r^4}.
\end{equation}
Using the expressions \eqref{GREB_sigma}$-$\eqref{LFEB_sigma} we obtain that, whenever $q^{2}_{_{\mathcal{E}}} = \sigma^{2} q^{2}_{_{\mathcal{B}}}$, the equations 
$G_{\alpha}{}^{\beta}\!- 8\pi(T_{\alpha}{}^{\beta})\!_{_{_{ E \! D}}}\!- 8\pi(T_{\alpha}{}^{\beta})\!_{_{_{D\!F}}} \! =0$ are fulfilled. 

Consider $\mathcal{H}(\mathcal{P})$ and $\mathcal{L}(\mathcal{F})$ related by the Legendre transform  
$\mathcal{H} = 2\mathcal{F}\mathcal{L}_{\mathcal{F}} -  \mathcal{L}$, with $\mathcal{H}_{\mathcal{P}} = 1/\mathcal{L}_{\mathcal{F}}$ and $\mathcal{P} = (\mathcal{L}_{\mathcal{F}})^{\!^{2}}\mathcal{F}$, then,  
the above theorem can be rewritten in terms of $(\mathcal{H},\mathcal{H}_{\mathcal{P}},\mathcal{P})$ in the following way.

{\bf Theorem II}, in the $(\mathcal{H},\mathcal{P})$-formalism: 
For any GR-$\mathcal{H}(\mathcal{P})$-DF theory with $\mathcal{H}(\mathcal{P})$ such that
\begin{equation}\label{HHPcons}
\mathcal{H}\bigg|_{_{\mathcal{P}=0}} = ~0  \quad\quad \textup{and}\quad\quad \mathcal{H}_{_{\mathcal{P}}}\bigg|_{_{\mathcal{P}=0}} =  ~ \frac{1}{\sigma},
\end{equation}
where $\big|_{_{\mathcal{P}=0}}$ means restricted to $\mathcal{P}=0$, and $\sigma$ is a nonzero positive constant, then the electrically and magnetically charged Ellis-Bronnikov WH metric \eqref{sigma_WH_BI_EB} together with a dust fluid with energy density \eqref{sigma_dust_BI_EB}, for the particular case $q^{2}_{_{\mathcal{E}}}=\sigma^{2}q^{2}_{_{\mathcal{B}}}$, is an exact solution 
of GR-$\mathcal{H}(\mathcal{P})$-DF equations \eqref{HPeqt}$-$\eqref{HPeqphi}, having
null invariant $\mathcal{P}$ 
\begin{equation}\label{Pcons}
\mathcal{P}\bigg|_{q^{2}_{_{\mathcal{E}}}=\sigma^{2}q^{2}_{_{\mathcal{B}}}} =  ~ 0. 
\end{equation}
{\bf Proof.} For an electromagnetic Hamiltonian density $\mathcal{H}(\mathcal{P})$ satisfying \eqref{HHPcons}, the non-null components of $(T_{\alpha}{}^{\beta})\!_{_{_{E\!D}}}$  
for an arbitrary static, spherically symmetric, electrically and magnetically charged spacetime metric having null electromagnetic invariant $\mathcal{P}=0$, are
\begin{equation}\label{HPEB_sigma}  
8\pi (T_{t}{}^{t})\!_{_{_{E\!D}}} = 8\pi (T_{r}{}^{r})\!_{_{_{E\!D}}}  = -\frac{2\sigma q^{2}_{_{\mathcal{B}}} }{ r^{4}},  \quad\quad 8\pi (T_{\theta}{}^{\theta})\!_{_{_{E\!D}}} = 8\pi (T_{\varphi}{}^{\varphi})\!_{_{_{E\!D}}} =  \frac{2q^{2}_{_{\mathcal{E}}}}{\sigma r^4}.
\end{equation}
Using the expressions \eqref{GREB_sigma}, \eqref{DFEB_sigma} and \eqref{HPEB_sigma} we obtain that, whenever $q^{2}_{_{\mathcal{E}}} = \sigma^{2} q^{2}_{_{\mathcal{B}}}$, the field equations, $G_{\alpha}{}^{\beta}\!- 8\pi(T_{\alpha}{}^{\beta})\!_{_{_{ E \! D}}}\!- 8\pi(T_{\alpha}{}^{\beta})\!_{_{_{D\!F}}} \! =0$, are fulfilled. 
\\
Alternatively, consider $\mathcal{H}(\mathcal{P})$ and $\mathcal{L}(\mathcal{F})$ related by the $\mathcal{F}\mathcal{P}$ duality  \eqref{FP_duality_tran}$-$\eqref{FP_duality_tranFuen}, one arrives at the following version of the theorem II: 
For any GR-$\mathcal{H}(\mathcal{P})$-DF theory with $\mathcal{H}(\mathcal{P})$ such that
\begin{equation}\label{HHPconsFP}
 \mathcal{H}\bigg|_{_{\mathcal{P}=0}} = ~ 0  \quad\quad \textup{and}\quad\quad \mathcal{H}_{_{\mathcal{P}}}\bigg|_{_{\mathcal{P}=0}} = ~ \tilde{\sigma}  
\end{equation}
where $\big|_{_{\mathcal{P}=0}}$ means restricted to $\mathcal{P}=0$, and $\tilde{\sigma}$ is a non-zero positive constant, then the electrically and magnetically charged Ellis-Bronnikov WH metric 
\begin{equation}\label{ssigma_WH_BI_EB}
\boldsymbol{ds}^{2} =  - \boldsymbol{dt}^{2} + \left( 1  -   \frac{q^{2}_{_{\mathcal{B}}} + \tilde{\sigma}^{2}q^{2}_{_{\mathcal{E}}}}{\tilde{\sigma}r^{2}}  \right)^{\!\!^{-1}} \boldsymbol{dr}^{2}
+ r^{2}(\boldsymbol{d\theta}^{2}  + \sin^{2}\theta \boldsymbol{d\varphi}^{2}), 
\end{equation}
together with a dust fluid with negative rest energy density given by 
\begin{equation}\label{ssigma_dust_BI_EB}   
\rho_{\!\!_{_{_{D\!F}}}}\!\!(r) = - \frac{q^{2}_{_{\mathcal{B}}} + \tilde{\sigma}^{2}q^{2}_{_{\mathcal{E}}}}{4\pi \tilde{\sigma} r^{4}} 
\end{equation}
for the particular case $q^{2}_{_{\mathcal{B}}}=\tilde{\sigma}^{2}q^{2}_{_{\mathcal{E}}}$, is an exact solution 
of GR-$\mathcal{H}(\mathcal{P})$-DF equations
having null invariant $\mathcal{P}$
\begin{equation}\label{PconsFP}
\mathcal{P}\bigg|_{q^{2}_{_{\mathcal{B}}}=\tilde{\sigma}^{2}q^{2}_{_{\mathcal{E}}}} =  ~ 0. 
\end{equation}
{\bf Proof.}
For the spacetime metric \eqref{ssigma_WH_BI_EB}, the Einstein tensor has the following nonzero components: 
\begin{equation}\label{GREB_ssigma}  
G_{t}{}^{t} = \frac{q^{2}_{_{\mathcal{B}}} + \tilde{\sigma}^{2}q^{2}_{_{\mathcal{E}}}}{ \tilde{\sigma} r^{4}}, \quad\quad G_{r}{}^{r} = - \frac{q^{2}_{_{\mathcal{B}}} + \tilde{\sigma}^{2}q^{2}_{_{\mathcal{E}}}}{ \tilde{\sigma} r^{4}}, \quad\quad G_{\theta}{}^{\theta} = G_{\varphi}{}^{\varphi}  = \frac{q^{2}_{_{\mathcal{B}}} + \tilde{\sigma}^{2}q^{2}_{_{\mathcal{E}}}}{ \tilde{\sigma} r^{4}}.  
\end{equation}
For an arbitrary static and spherically symmetric spacetime, the energy-momentum tensor of a dust fluid $(T_{\alpha}{}^{\beta})\!_{_{_{D\!F}}} = \rho_{\!\!_{_{_{D\!F}}}} u_{\alpha}u^{\beta}$ with rest energy density  $\rho_{\!\!_{_{_{D\!F}}}}$ given by \eqref{ssigma_dust_BI_EB} only has the following nonzero component:
\begin{equation}\label{DFEB_ssigma}  
8\pi (T_{t}{}^{t})\!_{_{_{D\!F}}} = - 8\pi\rho\!_{_{_{DF}}} = \frac{2q^{2}_{_{\mathcal{B}}} + 2\tilde{\sigma}^{2}q^{2}_{_{\mathcal{E}}}}{ \tilde{\sigma} r^{4}}.  
\end{equation}
On the other hand, in the $(\mathcal{H},\mathcal{P})$-formalism for a Hamiltonian density $\mathcal{H}(\mathcal{P})$ satisfying \eqref{HHPconsFP}, the non-null components $(T_{\alpha}{}^{\beta})\!_{_{_{E\!D}}}$ for an arbitrary static, spherically symmetric, electrically and magnetically charged spacetime metric having  
$\mathcal{P}=0$, are
\begin{equation}\label{HPEB_ssigma}  
{8\pi (T_{t}{}^{t})\!_{_{_{E\!D}}} = 8\pi (T_{r}{}^{r})\!_{_{_{E\!D}}}  = -\frac{2q^{2}_{_{\mathcal{B}}} }{ \tilde{\sigma} r^{4}},  \quad\quad 8\pi (T_{\theta}{}^{\theta})\!_{_{_{E\!D}}} = 8\pi (T_{\varphi}{}^{\varphi})\!_{_{_{E\!D}}} =  \frac{2\tilde{\sigma}q^{2}_{_{\mathcal{E}}}}{r^4}.}
\end{equation}
Using the expressions \eqref{GREB_ssigma}$-$\eqref{HPEB_ssigma}
we obtain that, whenever $q^{2}_{_{\mathcal{B}}} = \tilde{\sigma}^{2} q^{2}_{_{\mathcal{E}}}$, the Eqs. 
$G_{\alpha}{}^{\beta}\!- 8\pi(T_{\alpha}{}^{\beta})\!_{_{_{ E \! D}}}\!- 8\pi(T_{\alpha}{}^{\beta})\!_{_{_{D\!F}}} \! =0$ are fulfilled. 

Note that the methods \eqref{HHPcons}$-$\eqref{Pcons} and \eqref{HHPconsFP}$-$\eqref{PconsFP} are equivalent and the parameters $\sigma$ and $\tilde{\sigma}$ are related by $\sigma=1/\tilde{\sigma}$.

In this manner, we have established that the electrically and magnetically charged Ellis-Bronnikov WH metric \eqref{sigma_WH_BI_EB} together with a static and spherically symmetric dust fluid with negative rest energy density \eqref{sigma_dust_BI_EB}, for the particular case $q^{2}_{_{\mathcal{E}}}=\sigma^{2}q^{2}_{_{\mathcal{B}}}$, is a particular solution of GR-ED-DF equations for any electromagnetic Lagrangian density $\mathcal{L}(\mathcal{F})$ such that $\mathcal{L}(0)=0$ and $\mathcal{L}_{_{\mathcal{F}}}(0)=\sigma$  (or any electromagnetic Hamiltonian density $\mathcal{H}(\mathcal{P})$ such that $\mathcal{H}(0)=0$ and $\mathcal{H}_{_{\mathcal{P}}}(0)=\sigma^{-1}$).

For instance, as shown previously; since the Maxwell's electrodynamics \eqref{LED_Mx} satisfies \eqref{LLFcons} with $\sigma=1$, the electrically and magnetically charged Ellis-Bronnikov WH metric \eqref{EB_WH_EM} together with a static and spherically symmetric dust fluid with negative rest energy density \eqref{EB_WH_rho}, for the particular case $q^{2}_{_{\mathcal{E}}}=q^{2}_{_{\mathcal{B}}}$ is a solution of GR-$\mathcal{L}_{_{_{\mathrm{LED}}}}\!(\mathcal{F})$-DF equations; the Born-Infeld electrodynamics \eqref{BI} satisfy \eqref{LLFcons} with $\sigma=1$. We found that the electrically and magnetically charged Ellis-Bronnikov WH metric \eqref{WH_BI_EB} together with a static and spherically symmetric 
dust fluid with negative rest energy density \eqref{dust_BI_EB}, for the particular case $q^{2}_{_{\mathcal{E}}}=q^{2}_{_{\mathcal{B}}}$, is an exact solution of GR-$\mathcal{L}_{_{_{\mathrm{BI}}}}\!(\mathcal{F})$-DF equations.\\

\subsubsection{An example with an arbitrary parameter $\sigma$}
Consider the logarithmic electrodynamics model (LOG) which is defined by a Lagrangian density given by 
\begin{equation}\label{LogED}
\mathcal{L}_{_{_{\mathrm{LOG}}}} 
= \sigma \beta^{2}\ln\!\left( 1  + \frac{\mathcal{F}}{\beta^{2}}\right)     
\end{equation}
where $\sigma$ is a dimensionless constant that will be considered positive for successful phenomenology (fulfillment of WEC) and $\beta$ is a positive parameter that fixes the scale. This model satisfies 
\begin{equation}
\mathcal{L}_{_{_{\mathrm{LOG}}}}\bigg|_{_{_{\mathcal{F}=0}}} = ~ 0  \quad\quad\textup{and}\quad\quad 
\frac{d\mathcal{L}_{_{_{\mathrm{LOG}}}}}{d\mathcal{F}~~}\bigg|_{_{_{\mathcal{F}=0}}} = ~ \sigma.
\end{equation}
On another hand, according \eqref{F_invariant}, for this NLED theory the invariant $\mathcal{F}$ for a generic static and spherically symmetric spacetime geometry Eq.\!~\eqref{SSSmet} satisfies 
\begin{equation}\label{FdF}
\mathcal{F} =  \frac{q^{2}_{_{\mathcal{B}}}}{2r^{4}} - \frac{q^{2}_{_{\mathcal{E}}}}{2\sigma^{2}r^{4}}\!\!\left( 1 + \frac{\mathcal{F}}{\beta^{2}} \right)^{\!\!^{2}}
\end{equation}
which, for the general case $(q_{_{\mathcal{E}}}\neq0, ~\forall q_{_{\mathcal{B}}})$, the solution\footnote{
The other solution of Eq.\!~\eqref{FdF}, for the general case $(q_{_{\mathcal{E}}}\neq0, ~\forall q_{_{\mathcal{B}}})$, is
\begin{equation}
\mathcal{F} = -\frac{\beta^{2}\left(q^{2}_{_{\mathcal{E}}} + \sigma^{2}\beta^{2}r^{4}\right)}{q^{2}_{_{\mathcal{E}}}} \left[ 1 + \sqrt{1 + \frac{ \sigma^{2} q^{2}_{_{\mathcal{E}}}q^{2}_{_{\mathcal{B}}} - q^{4}_{_{\mathcal{E}}} }{ \beta^{4}r^{8}\left( \sigma^{2} + \frac{q^{2}_{_{\mathcal{E}}}}{\beta^{2}r^{4}} \right)^{\!\!^{2}} } } \right] 
\end{equation}
which we discard since they diverge when $r\rightarrow\infty$ [in contrast to \eqref{F_log_NLED} which tends to zero as $r$ tends to infinity] and this could be viewed as incompatible with the assumption of asymptotically flat spacetime geometry. } for $\mathcal{F}$ is
\begin{equation}\label{F_log_NLED}
\mathcal{F} = -\frac{\beta^{2}\left(q^{2}_{_{\mathcal{E}}} + \sigma^{2}\beta^{2}r^{4}\right)}{q^{2}_{_{\mathcal{E}}}} \!\!\left[ 1 - \sqrt{1 + \frac{ \sigma^{2} q^{2}_{_{\mathcal{E}}}q^{2}_{_{\mathcal{B}}} - q^{4}_{_{\mathcal{E}}} }{ \beta^{4}r^{8}\left( \sigma^{2} + \frac{q^{2}_{_{\mathcal{E}}}}{\beta^{2}r^{4}} \right)^{\!\!^{2}} } } \right]
\end{equation}
and being such that $\mathcal{F}\big|_{q^{2}_{_{\mathcal{E}}} = \sigma^{2} q^{2}_{_{\mathcal{B}}}} = ~ 0$.
Indicating, according to the previous theorem, that the electrically and magnetically charged
Ellis-Bronnikov wormhole metric \eqref{sigma_WH_BI_EB} together with a dust fluid with negative energy density \eqref{sigma_dust_BI_EB}, for the particular case $q^{2}_{_{\mathcal{E}}} = \sigma^{2} q^{2}_{_{\mathcal{B}}}$ is a particular solution of GR-$\mathcal{L}_{_{_{\mathrm{LOG}}}}\!(\mathcal{F})$-DF. 

\begin{itemize}
\item{\bf Most general ultrastatic and spherically symmetric solution of GR-$\boldsymbol{\mathcal{L}_{_{_{\mathrm{LOG}}}}\!(\mathcal{F})}$-DF with $\boldsymbol{(q_{_{\mathcal{E}}}\neq0, ~\forall q_{_{\mathcal{B}}})}$} 

For the case $(q_{_{\mathcal{E}}}\neq0, ~\forall q_{_{\mathcal{B}}})$, substituting \eqref{LogED} into the Eqs. \eqref{UltraS_SolutionEM}$-$\eqref{density_dust_EM_LF} we arrive at the following metric
\begin{equation}\label{TWH_log_EM}
\boldsymbol{ds}^{2} = - ~ \boldsymbol{dt}^{2}  + \left[\! 1 \!- 2\sigma\beta^{2}r^{2} \ln\!\!\left( 1 + \frac{\mathcal{F}}{\beta^{2}} \right) - \frac{2q^{2}_{_{\mathcal{E}}}}{\sigma r^{2}}\!\!\left( 1 + \frac{\mathcal{F}}{\beta^{2}} \right) \!\right]^{\!\!^{-1}}\!\!\boldsymbol{dr}^{2} + r^{2}(\boldsymbol{d\theta}^{2} + \sin^{2}\theta \boldsymbol{d\varphi}^{2}) 
\end{equation}
where $\mathcal{F}$ is given by \eqref{F_log_NLED}, 
and a dust fluid with energy density given by,
\begin{equation}\label{rho_log_EM} 
\rho_{\!\!_{_{_{D\!F}}}}\!\!(r)\!=\!  
\frac{\sigma\beta^{2}\!\ln\!\!\left[\! 1 \!- \frac{\sigma^{2}\beta^{2}r^{4} \!+ q^{2}_{_{\mathcal{E}}}}{q^{2}_{_{\mathcal{E}}}}\!\!\left(\!\! 1 \!- \!\!\sqrt{1 \!+ \frac{  \sigma^{2} q^{2}_{_{\mathcal{E}}}q^{2}_{_{\mathcal{B}}} \! - q^{4}_{_{\mathcal{E}}} }{ \beta^{4}r^{8}\left( \sigma^{2} + \frac{q^{2}_{_{\mathcal{E}}}}{\beta^{2}r^{4}} \right)^{\!\!^{2}} } }
\!\right)\!\!\right]}{2\pi}\!  
- \frac{\sigma q^{2}_{_{\mathcal{B}}} \!\!\left[\! 1 \!- \!\frac{\sigma^{2}\beta^{2}r^{4} \! + q^{2}_{_{\mathcal{E}}}}{q^{2}_{_{\mathcal{E}}}}\!\!\left(\!\! 1 \!- \!\sqrt{1 \!+ \frac{  \sigma^{2}q^{2}_{_{\mathcal{E}}}q^{2}_{_{\mathcal{B}}} - q^{4}_{_{\mathcal{E}}} }{ \beta^{4}r^{8}\!\left(\! \sigma^{2} \!+ \frac{q^{2}_{_{\mathcal{E}}}}{\beta^{2}r^{4}}\!\right)^{\!\!^{2}} }}\right)\!\!\right]^{\!\!^{-1}} 
}{2\pi r^{4}}.
\end{equation}
It is important to emphasize that for the particular case
$q^{2}_{_{\mathcal{E}}} = \sigma^{2}q^{2}_{_{\mathcal{B}}}$ [the electromagnetic invariant  \eqref{F_log_NLED}, becomes zero $\mathcal{F}\big|_{q^{2}_{_{\mathcal{E}}} = \sigma^{2} q^{2}_{_{\mathcal{B}}}} = ~ 0$], whereas the Eqs.\!~\eqref{TWH_log_EM} and \eqref{rho_log_EM} respectively become the electrically and magnetically charged Ellis-Bronnikov wormhole metric \eqref{sigma_WH_BI_EB} and
dust energy density \eqref{sigma_dust_BI_EB}, being in agreement with the previous theorem. 

\item {\bf Purely magnetic, ultrastatic and spherically symmetric solution of GR-$\boldsymbol{\mathcal{L}_{_{_{\mathrm{LOG}}}}\!(\mathcal{F})}$-DF}
 
For the purely magnetic case ($q^{2}_{_{\mathcal{E}}} =0,~q^{2}_{_{\mathcal{B}}}=q$), the Eq.\!~\eqref{FdF} reduces to
$\mathcal{F} =  \frac{q^{2}}{2r^{4}}$, and evaluating \eqref{UltraS_Solution}$-$\eqref{density_dust} yields the following purely magnetic, ultrastatic and spherically symmetric solution of GR-$\mathcal{L}_{_{_{\mathrm{LOG}}}}\!(\mathcal{F})$-DF, characterized by the spacetime metric 
\begin{equation}\label{TWH_log_M}
\boldsymbol{ds}^{2} = - ~ \boldsymbol{dt}^{2}  + \left[ 1 - 2\sigma\beta^{2} r^{2} \ln\left( 1 + \frac{q^{2}}{2\beta^{2}r^{4}}\right) \!\right]^{\!\!^{-1}}\!\!\boldsymbol{dr}^{2}  + ~ r^{2}(\boldsymbol{d\theta}^{2} + \sin^{2}\theta \boldsymbol{d\varphi}^{2})
\end{equation}      
 and a dust fluid with energy density 
\begin{equation}\label{rho_log_M}
\rho_{\!\!_{_{_{D\!F}}}}\!\!(r) = \frac{\sigma\beta^{2}}{2\pi}\!\!\left[ \ln\left( 1 + \frac{q^{2}}{2\beta^{2}r^{4}}\right) - \frac{2q^{2}}{2\beta^{2}r^{4} + q^{2} } \right].
\end{equation}
In this solution, the metric component $g^{rr}$ belongs to the class $\mathcal{C}^{\infty}$ for all $r$,    
furthermore, it satisfies $\lim\limits_{r\to0}g^{rr} = \lim\limits_{r\to\infty}g^{rr} = 1$, and in the region $r\in(0,\infty)$ has  an absolute minimum ($r=r_{_{min}}$) such that $g^{rr}(r_{_{min}})$ can be negative, positive or null. Specifically, for the case $g^{rr}(r_{_{min}})<0$, the solution describes a wormhole spacetime geometry without event horizon. On the other hand, for this solution, the conditions \eqref{NEC_NLED}$-$\eqref{WEC_NLED} are trivially satisfied.
Whereas, for the wormhole configuration, the dust energy density \eqref{rho_log_M} is negative defined, hence the NEC and WEC are violated, Eq.\!~\eqref{NEC_SF}.

Finally, we can mentioned that, since the electromagnetic solution \eqref{TWH_log_EM}$-$\eqref{rho_log_EM} is not defined by $q_{_{\mathcal{E}}}=0$, then it is not connected with the purely magnetic solution \eqref{TWH_log_M}$-$\eqref{rho_log_M}.

\end{itemize}


\section{CONCLUDING REMARKS}\label{secV}

Static and spherically symmetric spacetime metrics, with $\Phi(r)\neq constant$, are not possible within the framework of GR-ED-DF. Nevertheless, those with $\Phi(r)=constant$, are not excluded.    

This study concludes that in the context of general relativity, ultrastatic and spherically symmetric spacetimes (i.e., static and spherically symmetric solutions with constant redshift functions) supported by (linear or nonlinear) electromagnetic fields and dust fluids lead to the formation of numerous Morris-Thorne wormholes.\\
To accomplish this, we have presented a procedure to obtain ultrastatic, spherically symmetric, electrically and magnetically charged spacetime solutions of the GR-ED-DF gravity
in the $(\mathcal{L},\mathcal{F})$ and $(\mathcal{H},\mathcal{P})$ formalisms.
Within the $(\mathcal{L},\mathcal{F})$ formalism, starting from an arbitrary electromagnetic Lagrangian density
$\mathcal{L}(\mathcal{F})$  our method determines the energy density of a dust fluid (that is $2\pi \rho_{\!\!_{_{_{D\!F}}}} = \mathcal{L} - F_{\theta\varphi}F^{\theta\varphi}\mathcal{L}_{\mathcal{F}}$) in order that the electrically and magnetically charged spacetime metric Eq.\!~\eqref{UltraS_SolutionEM} becomes the more general ultrastatic and spherically symmetric solution of the GR-$\mathcal{L}(\mathcal{F})$-DF field equations. 
While, in the $(\mathcal{H},\mathcal{P})$ formalism, starting from an arbitrary electromagnetic Hailtonian density $\mathcal{H}(\mathcal{P})$  our method determines the energy density of a dust fluid (that is $2\pi \rho_{\!\!_{_{_{D\!F}}}} = P_{tr}P^{tr}\mathcal{H}_{\mathcal{P}} - \mathcal{H}$) in order that the electrically and magnetically charged spacetime metric Eq.\!~\eqref{UltraS_SolutionEM_HP} becomes the more general ultrastatic and spherically symmetric solution of the GR-$\mathcal{H}(\mathcal{P})$-DF field equations.
We provide some examples using several physically relevant electrodynamics models. 
Using Maxwell's theory of electrodynamics (linear electrodynamics)   
$\mathcal{L}_{_{_{\mathrm{LED}}}} \!=\! \mathcal{F}$, we obtain  
a solution which can be interpreted as an  
electrically and magnetically charged Ellis-Bronnikov wormhole.  
The nonlinear electrodynamics (NLED) models;
Born-Infeld (BI), $\mathcal{L}_{_{_{\mathrm{BI}}}} \!=\! -4\beta^{2} + 4\beta^{2} \sqrt{ 1 + \mathcal{F}\!/\!(2\beta^{2})~}$; Euler-Heisenberg (EH) in the approximation of the weak-field limit, $\mathcal{L}_{_{_{\mathrm{EH}}}} \!=\! \mathcal{L}_{_{_{\mathrm{LED}}}} - \mu^{2} \mathcal{F}^{2}\!/2$; and logarithmic electrodynamics (LOG), 
$\mathcal{L}_{_{_{LOG}}} = \sigma \beta^{2}\ln\!\left( 1 + \frac{\mathcal{F}}{\beta^{2}}\right)$, are also used. 
With those NLED models, in this work, three new electromagnetically 
charged traversable wormholes (BI-DF, EH-DF and LOG-DF wormholes) are presented as exact solutions in the context of general relativity. 
We show that in the limit of the weak electromagnetic field, each of our solutions become a traversable wormhole of the electromagnetically charged Ellis-Bronnikov wormhole type.\\
In each of our T-WH solutions, the exotic dust fluid is solely responsible for the violation of the null-energy condition required for the traversability of wormhole, see Eq.\!~\eqref{NEC_SF}. Whereas, according to Eqs.\!~\eqref{UltraS_SolutionEM} and \eqref{UltraS_SolutionEM_HP}, the T-WH geometry cannot have a nonzero  shape function without a linear (or nonlinear) electrodynamics source. Nevertheless, the results from Eqs.\!~\eqref{density_dust_EM_LF} and \eqref{density_dust_EM_HP}, which imply that the energy density of the dust fluid originates from the electrodynamics model, suggest that the electromagnetic source can minimize the amount of exotic dust fluid in the T-WH. \\
Additionally, we prove a theorem that determines when a GR-$\mathcal{L}(\mathcal{F})$-DF theory admits an electrically and magnetically charged Ellis-Bronnikov wormhole spacetime metric (with vanishing electromagnetic invariant $\mathcal{F}$) as a particular solution. This theorem can be mapped to the $(\mathcal{H},\mathcal{P})-$formalism by two different methods, through a Legendre transform and through  $\mathcal{F}\mathcal{P}$ duality, and it is found that  the results of both mappings are equivalent. \\
Therefore, we can conclude that, within the framework of GR-ED-DF, the only permissible static and spherically symmetric T-WHs are exclusively the ultrastatic and spherically symmetric types, meaning those static and spherically symmetric T-WHs with a trivial redshift function, i.e., with $\Phi(r)=constant$. \\ 
In a forthcoming paper we shall explore the stability of T-WHs obtained in this work.  
Furthermore, in view of the method \eqref{UltraS_SolutionEM}$-$\eqref{density_dust_EM_LF} for the $(\mathcal{L},\mathcal{F})-$formalism [or \eqref{UltraS_SolutionEM_HP}$-$\eqref{density_dust_EM_HP} for the $(\mathcal{H},\mathcal{P})-$formalism] presented in this work, 
it may also be interesting to investigate whether the stability of the electromagnetically charged wormhole can be guaranteed by a suitable choice of $\mathcal{L}(\mathcal{F})$ or $\mathcal{H}(\mathcal{P})$. 

\appendix

\section{NULL AND WEAK ENERGY CONDITIONS IN GR-ED-DF}

Next, within the GR-ED-DF context, we show the conditions under which the matter sources of the generic metric \eqref{SSSmet}  satisfy the null and weak energy conditions everywhere. 

For an energy-momentum tensor $T_{\mu\nu}$, the null energy condition (NEC) stipulates that for every null vector, $n^{\alpha}$, yields $T_{\mu\nu}n^{\mu}n^{\nu}\geq0$.
Following \cite{WEC}, for a diagonal energy-momentum tensor $(T_{\alpha\beta})=\mathrm{diag} \left( T_{tt},T_{rr},T_{\theta\theta},T_{\varphi\varphi} \right)$, which can be
conveniently written as, 
\begin{equation}\label{diagonalEab}
T_{\alpha}{}^{\beta} = - \rho_{t} \hskip.05cm \delta_{\alpha}{}^{t}\delta_{t}{}^{\beta} + P_{r} \hskip.05cm \delta_{\alpha}{}^{r}\delta_{r}{}^{\beta} + P_{\theta} \hskip.05cm \delta_{\alpha}{}^{\theta}\delta_{\theta}{}^{\beta} + P_{\varphi} \hskip.05cm \delta_{\alpha}{}^{\varphi}\delta_{\varphi}{}^{\beta} ,
\end{equation}
 where $\rho_{t}$ may be interpreted as the rest energy density of the matter,  
 whereas $P_{r}$, $P_{\theta}$ and $P_{\varphi}$ are respectively the pressures along the $r$, $\theta$ and $\varphi$ directions. In terms of (\ref{diagonalEab}) the NEC implies
\begin{equation}\label{NEC_Eab}
\rho_{t} + P_{a} \geq0 \quad\quad\textup{with}\quad\quad  a = \{ r, \theta, \varphi \}.
\end{equation}
The weak energy condition (WEC) states that for any timelike vector $\boldsymbol{k} = k^{\mu}\partial_{\mu}$, ({\it i.e.,} $k_{\mu}k^{\mu}<0$), the energy-momentum tensor  obeys the inequality
$T_{\mu\nu}k^{\mu}k^{\nu} \geq 0$, which means that the local energy density $\rho_{\!_{_{loc}}}= T_{\mu\nu}k^{\mu}k^{\nu}$ as measured by any observer with timelike vector 
$\boldsymbol{k}$ is a non-negative quantity.  
For an energy-momentum tensor of the form (\ref{diagonalEab}), the WEC will be satisfied if and only if,
\begin{equation}\label{WEC}
\rho_{t} = - T_{t}{}^{t} \geq0, \quad\quad\quad\quad \rho_{t} + P_{a} \geq0 \quad\textup{with}\quad  a = \{ r, \theta, \varphi \}.
\end{equation}
Therefore, according to the Eqs. \eqref{NEC_Eab} and \eqref{WEC}, if NEC is violated, then WEC is violated as well. 

\begin{itemize}
\item {\bf NEC and WEC for a dust fluid}

Identifying \eqref{T_DF} with \eqref{diagonalEab} yields,
\begin{equation}
(\rho_{t})\!_{_{_{D\!F}}} = \rho_{\!\!_{_{_{D\!F}}}}, \quad\quad\quad\quad 
(P_{r})\!_{_{_{D\!F}}} = (P_{\theta})\!_{_{_{D\!F}}} = (P_{\varphi})\!_{_{_{D\!F}}} = 0      
\end{equation}
since $(P_{a})\!_{_{_{D\!F}}}=0$ for all $a=r$, $\theta$, $\varphi$, the energy-momentum tensor $(T_{\alpha}{}^{\beta})\!_{_{_{D\!F}}}$ satisfies the NEC, as well as satisfies the WEC, if  
\begin{equation}\label{NEC_SF}
\rho_{\!\!_{_{_{D\!F}}}} \geq 0.
\end{equation}
\item {\bf NEC and WEC for the linear/nonlinear electrodynamics field in the $(\mathcal{L},\mathcal{F})-$formalism 
}

Identifying \eqref{TEM_ED} with \eqref{diagonalEab}, and using Eqs.\!~\eqref{Fab_SSS}$-$\eqref{F_invariant}, 
yields,
\begin{eqnarray}
8\pi (\rho_{t})\!_{_{_{E\!D}}} = - 8\pi (P_{r})\!_{_{_{E\!D}}} =  \frac{ 2q^{2}_{_{\mathcal{E}}}  }{r^{4}\mathcal{L}_{\mathcal{F}}} + 2\mathcal{L},
\quad\quad\quad 8\pi (P_{\theta})\!_{_{_{E\!D}}} = 8\pi (P_{\varphi})\!_{_{_{E\!D}}} =  \frac{2q^{2}_{_{\mathcal{B}}}}{r^4}\mathcal{L}_{\mathcal{F}} - 2\mathcal{L}  
\end{eqnarray}
since $(\rho_{t})\!_{_{_{E\!D}}} + (P_{r})\!_{_{_{E\!D}}}=0$, and 
\begin{equation}\label{NEC_NLEDa}
8\pi(\rho_{t})\!_{_{_{E\!D}}} + 8\pi (\rho_{\theta})\!_{_{_{E\!D}}} = 8\pi(\rho_{t})\!_{_{_{E\!D}}} + 8\pi(\rho_{\varphi})\!_{_{_{E\!D}}}  = \frac{2}{r^{4}\mathcal{L}_{\mathcal{F}}}\left(q^{2}_{_{\mathcal{E}}} + q^{2}_{_{\mathcal{B}}}\mathcal{L}^{2}_{\mathcal{F}}\right) 
\end{equation}
the tensor $(T_{\alpha}{}^{\beta})\!_{_{_{E\!D}}}$ satisfies the NEC if, 
\begin{equation}\label{NEC_NLED}
\mathcal{L}_{\mathcal{F}}\geq0.
\end{equation}
in addition to (\ref{NEC_NLED}) if,
\begin{equation}\label{WEC_NLED}
8\pi (\rho_{t})\!_{_{_{E\!D}}} = \frac{ 2q^{2}_{_{\mathcal{E}}}  }{r^{4}\mathcal{L}_{\mathcal{F}}} + 2\mathcal{L}\geq0, 
\end{equation}
the WEC is satisfied.
\item {\bf NEC and WEC for the linear/nonlinear electrodynamics field in the $(\mathcal{H},\mathcal{P})-$formalism }

Identifying \eqref{HP_EM_tensr} with \eqref{diagonalEab}, for static and spherically symmetric configurations \eqref{P_invariant},  yields
%
\begin{eqnarray}
8\pi (\rho_{t})\!_{_{_{E\!D}}} = - 8\pi (P_{r})\!_{_{_{E\!D}}} =  \frac{ 2q^{2}_{_{\mathcal{B}}} }{r^{4}\mathcal{H}_{\mathcal{P}}} - 2\mathcal{H},
\quad\quad\quad 8\pi (P_{\theta})\!_{_{_{E\!D}}} = 8\pi (P_{\varphi})\!_{_{_{E\!D}}} =  
\frac{2q^{2}_{_{\mathcal{E}}}}{r^4}\mathcal{H}_{\mathcal{P}} + 2\mathcal{H}
\end{eqnarray}
since $(\rho_{t})\!_{_{_{E\!D}}} + (P_{r})\!_{_{_{E\!D}}}=0$, and 
\begin{equation}\label{NEC_NLEDaHP}
8\pi(\rho_{t})\!_{_{_{E\!D}}} + 8\pi (\rho_{\theta})\!_{_{_{E\!D}}} = 8\pi(\rho_{t})\!_{_{_{E\!D}}} + 8\pi(\rho_{\varphi})\!_{_{_{E\!D}}}  = \frac{2}{r^{4}\mathcal{H}_{\mathcal{P}}}\left(q^{2}_{_{\mathcal{B}}} + q^{2}_{_{\mathcal{E}}}\mathcal{H}^{2}_{\mathcal{P}}\right) 
\end{equation}
the tensor $(T_{\alpha}{}^{\beta})\!_{_{_{E\!D}}}$ satisfies the NEC if, 
\begin{equation}\label{NEC_NLEDHP}
\mathcal{H}_{\mathcal{P}}\geq0.
\end{equation}
in addition to (\ref{NEC_NLEDHP}) if,
\begin{equation}\label{WEC_NLEDHP}
8\pi (\rho_{t})\!_{_{_{E\!D}}} =  \frac{ 2q^{2}_{_{\mathcal{B}}} }{r^{4}\mathcal{H}_{\mathcal{P}}} - 2\mathcal{H}    \geq 0
\end{equation} 
the WEC is satisfied.
\end{itemize}

\end{document}